\documentclass{article}
\pdfoutput=1
\usepackage[a4paper]{geometry}
\usepackage{graphicx} %
\usepackage{amssymb,amsmath}
\usepackage{tikz,tikz-feynman}
\usepackage{siunitx}

\usepackage{hyperref}

\newcommand{\dd}{\mathrm{d}}
\newcommand{\cross}{\times}
\newcommand{\R}{\mathbb{R}}
\newcommand{\bb}[1]{\boldsymbol{#1}}
\newcommand{\tr}{\mathrm{tr}}

\usepackage{xcolor}

\definecolor{red}{rgb}{1,0,0}

 \definecolor{darkgreen}{rgb}{0, .7, 0}

 \definecolor{purple}{rgb}{.7, 0, 1}

\definecolor{col1}{RGB}{100,143,255}
\definecolor{col2}{RGB}{120, 94, 240}
\definecolor{col3}{RGB}{254,97,0}
\definecolor{col4}{RGB}{220, 38, 127}
\definecolor{col5}{RGB}{255, 176, 0}

\hypersetup{
  pdfauthor={Michael Borinsky, Mathijs Fraaije},
  pdftitle={Tropical sampling from Feynman measures}
  linkcolor  = col2,
  citecolor  = col4,
  urlcolor   = col1,
  colorlinks = true,
}

\tikzfeynmanset{warn luatex=false}
\begin{document}
\title{Tropical sampling from Feynman measures}

\date{}

\author{Michael Borinsky%
\thanks{Perimeter Institute, 31 Caroline St N, Waterloo, ON N2L 2Y5, Canada}%
\and Mathijs Fraaije%
\thanks{Universität Bern, Institute for Theoretical Physics, Gesellschaftsstrasse 2, 3012 Bern, Switzerland}
} %
\maketitle
\begin{abstract}
We introduce an algorithm that samples a set of loop momenta distributed as a given Feynman integrand. The algorithm uses the tropical sampling method and can be applied to evaluate phase-space-type integrals efficiently. We provide an implementation, \texttt{momtrop}, and apply it to a series of relevant integrals from the loop-tree duality framework. Compared to naive sampling methods, we observe convergence speedups by factors of more than $10^6$.
\end{abstract}
\section{Introduction}
Feynman integrals are a key tool in obtaining accurate predictions from quantum field theories.
In the loop representation, given a Feynman graph~$G$, these integrals are of the shape,
\begin{align} \label{eq:feyn} I_G = \int \frac{\prod_{\ell=1}^L \dd^{D}{{k}_{\ell}} }{\prod_{e=1}^E D_e(\bb k)^{\nu_e}} \;, \end{align}
where the product over $\ell$ runs over all $L$ \emph{loops} of $G$ and the one over $e$ runs over all $E$ \emph{edges (also called propagators, or lines)} of $G$, the \emph{propagators} $D_e$ are quadratic functions in the loop momenta $k_\ell$, the external momenta $p_i$, and the masses $m_e$, the $\nu_e$ are arbitrary propagator weights, and we integrate over a copy of $\R^D$ for each loop. Throughout the paper, we will omit the dependence of quantities such as $I_G$ and $D_e(\bb k)$ on the external momenta $\bb p = (p_1, p_2, \ldots)$ and the masses $\bb m= (m_1,m_2,\ldots)$ to keep the notation lean.

We will introduce an algorithm that efficiently produces samples from the associated \emph{Feynman measure} in the space of all $L$ loop momenta $\bb k = (k_1,\ldots,k_L) \in (\R^{D})^L$:
\begin{align} \label{eq:mu} \mu_G = \frac{1}{I_G} \frac{\prod_{\ell=1}^L \dd^{D}{{k}_{\ell}} }{\prod_{e=1}^E D_e(\bb k)^{\nu_e}} \;. \end{align}
If we work in Minkowski spacetime, this density over loop momentum space $(\R^{D})^L$ is, in general, not smooth, and the propagators $D_e$ are only well-defined thanks to the $i\varepsilon$ prescription. Moreover, they are inherently complex-valued. It is unclear how to sample from such general distributions. Here, we focus on (at least effectively) Euclidean spacetimes. In this specific case, each propagator in the density~\eqref{eq:mu} is real and positive. For $\mu_G$ to be a well-defined density over loop-momentum space, we further require that $I_G$ has a finite value. This way, $\mu_G$ is normalized to $1$ as a density over $(\R^D)^L$, and positivity guarantees that we may interpret $\mu_G$ as a probability measure. Explicitly, we require the propagators to be of the form:
\begin{align} \label{eq:prop} D_e(\bb k) = \left(\sum_{\ell=1}^{L} \mathcal{M}_{e, \ell} k_\ell + p_e \right)^2 + m_e^2 \;, \end{align}
where $\mathcal M$ is an $E\times L$ matrix that is fixed by the topology of the graph and a choice of a loop routing through it, $p_e$ and $m_e$ are the external momentum flowing through $e$ and the mass of the propagator $e$, respectively, and the vector square is taken using the all-plus-sign Euclidean metric. Further, we assume all the propagator weights $\nu_e$ are positive real numbers. Note that we encode the external momenta in a slightly unusual form: $p_e$ denotes the external momentum flowing through the edge $e$. 
After fixing a momentum routing through the graph, $p_e$ is given by a linear combination of the external momenta flowing into the legs of the graph.
We describe an algorithm that efficiently samples from the resulting measure~\eqref{eq:mu} in Section~\ref{sec:algo}. We implemented this algorithm as the \texttt{Rust} software package \texttt{momtrop} that can be conveniently used for Feynman or phase-space integration problems. This implementation is discussed in Section~\ref{sec:imp}.

An interesting application of our sampling algorithm for the Feynman measure $\mu_G$ lies in collider physics phenomenology. Making a high-accuracy prediction in high-energy physics can be (roughly) seen as a two-step process: First, computing the required loop integrals, and second, integrating the resulting $S$-matrix element over a phase space of possible measurements. We argue that both steps can be accelerated significantly using dedicated sampling algorithms for the Feynman measure~\eqref{eq:mu}. The reason for this is that many integration problems of Feynman integrals or phase space integrals in four-dimensional Minkowski space  can be put into the form:
\begin{align} \label{eq:kernel} \int_{(\R^{3})^L} f(\bb k) \cdot \mu_G, \end{align}
where we integrate over $L$ copies of three-dimensional Euclidean space, and the residual integration kernel $f(\bb k)$ is a (possibly quite complicated) function that depends on all the momenta $\bb k = (k_1,\ldots,k_L)$ but has fewer (integrable) singularities in the integration domain than the Feynman integrand~\eqref{eq:feyn}. In this way, the integrable singularities of the integrands can be `absorbed into the measure $\mu_G$.' The integral~\eqref{eq:kernel} can then be evaluated by sampling the Feynman measure $\mu_G$ and repeatedly evaluating the kernel $f(\bb k)$ within a typical Monte Carlo workflow. The increased regularity of the kernel $f(\bb k)$ will either make an integral numerically integrable or increase the convergence rate significantly.

An alternative viewpoint is that our sampling algorithm for the Feynman measure~\eqref{eq:mu} provides an efficient way to perform \emph{importance sampling} over a complicated phase space measure. %
In Section~\ref{sec:visual}, we provide visualizations of our sampling algorithm concerning this viewpoint, showing how our algorithm increases the sampling density near (integrable) singularities of the integrand.

Early algorithms for sampling over phase space volumes did not consider the singularity structure of the $S$-matrix. For example, the \texttt{RAMBO} sampling strategy~\cite{Kleiss:1985gy} indiscriminately generates points from phase space. The phase space generator \texttt{PHEGAS} \cite{Papadopoulos:2000tt} takes a more sophisticated approach, accounting for structures such as Breit-Wigner peaks. Here in the present paper, we introduce a new sampling strategy that systematically focuses on more relevant regions of phase space in a way that is naturally guided by the deep (tropical) geometric structure of the Feynman measure. Harnessing this geometric structure enables a severe reduction of the overall computational costs.

 Binoth, Gehrmann-De Ridder, Gehrmann, and Heinrich used sector decomposition techniques to deal with integrable singularities in phase-space integrals \cite{Gehrmann-DeRidder:2003pne,Binoth:2004jv}. Parallel to their approach, we use the tropical sampling method \cite{Borinsky:2020rqs,Borinsky:2023jdv}, which draws ideas from sector decomposition \cite{Binoth:2000ps,Kaneko:2009qx} and from the works of Panzer \cite{Panzer:2019yxl} and Brown \cite{Brown:2015fyf}, to achieve our goal of sampling from non-uniform Feynman-type measures. In \ref{ap:example}, we give a detailed, didactic example that illustrates the  tropical sampling strategy.

A particularly well-suited domain of application for our algorithm is the loop-tree duality framework~\cite{Catani:2008xa,Bierenbaum:2010cy}. We discuss the application of our software library \texttt{momtrop} to LTD problems in Section~\ref{sec:ltd}. This discussion illustrates the substantial performance improvements that can be achieved using our refined sampling approach. We provide benchmarks of the runtime and the accuracy that compare the new method with naive sampling methods. 
For practical applications, such naive sampling methods are usually combined with variance reduction methods such as the VEGAS algorithm~\cite{Lepage:1977sw}, \emph{multi-channeling}, or carefully designed choices of phase-space measures. As these variance reduction methods come with many free parameters whose optimization for the specific problem under inspection is a nontrivial task, we do not attempt to broadly compare our new approach with such general methods. 
Moreover, these techniques typically require readjustment for each new phase-space point, adding to the computational overhead.  A key advantage of the tropical approach is that the preprocessing step (at loop orders below $\approx 8$) has negligible computational cost and is independent of the specific values of masses and external momenta; therefore, a single preprocessing step suffices to handle large regimes of parameter space efficiently.
Exemplary, in Section~\ref{sec:multi}, we discuss the (still substantial) performance improvements of our method in comparison to the \emph{multi-channeling}~\cite{Kleiss:1994qy} variance reduction method.  We conclude in Section~\ref{sec:con}.
\section{Algorithm to sample from the Feynman measure}
\label{sec:algo}
\subsection{Overview}
\label{sec:overview}

We will explain the algorithm based on a more conveniently normalized version of the Feynman integral~\eqref{eq:kernel} and show  how to evaluate via Monte Carlo sampling, the general integral
\begin{align} \label{eq:int1} I_{G,f}= \int_{(\R^{D})^L} \frac{f(\bb k)}{\prod_{e=1}^E D_e(\bb k)^{\nu_e}} \frac{\dd^{D}{{k}_{1}} }{\pi^{D/2}} \cdots \frac{\dd^{D}{{k}_{L}} }{\pi^{D/2}} , \end{align}
where $D_e$ is given as in Eq.~\eqref{eq:prop} with masses and Euclidean external momenta encoded in a Feynman graph $G$, the $\nu_e$ are real numbers $>0$, and
$f(\bb k)$ is any function in the loop momenta $\bb k=(k_1,\ldots,k_L)$. Additionally, convergence has to be ensured for $f(\bb k)=1.$

Instead of directly producing a sample from the Feynman measure~\eqref{eq:mu}, our algorithm will produce samples $(\bb k, W)$ from the related \emph{weighted Feynman measure} $\widetilde \mu_G$ on the augmented space $(\R^{D})^L \times [0,W_{\mathrm{max}}]$, where $W$ is a weight factor confined to an interval $0 \leq W \leq W_{\mathrm{max}}$. Samples from the unweighted Feynman measure $\mu_G$ can be obtained via an additional \emph{rejection-sampling} step, which is feasible due to the bounded weight. For many typical integration problems, however, this step is not necessary.

The weighted Feynman measure $\widetilde \mu_G$ is conveniently described by using it to rewrite the integral~\eqref{eq:int1}:
\begin{align} \label{eq:int2} I_{G,f}= Z_G \cdot \int_{(\R^{D})^L \times [0,W_{\mathrm{max}}]} W\cdot f(\bb k) \cdot \widetilde \mu_G , \end{align}
where $Z_G$ is a prefactor computed in a preprocessing step of the algorithm. After this preprocessing step, our algorithm efficiently produces samples $(\bb k, W) \in (\R^{D})^L \times [0,W_{\mathrm{max}}]$ distributed according to the  
weighted Feynman measure $\widetilde \mu_G$. So, assuming convergence, we may use a standard Monte Carlo approach to approximate $I_{G,f}$ via
\begin{align} \label{eq:mc} I_{G,f}= \frac{Z_G}{N} \cdot \sum_{i =1}^N\; W_i \cdot f(\bb k^{(i)}) + \mathcal O\left(\frac{1}{\sqrt{N}}\right) , \end{align}
where $(\bb k^{(1)},W_1), \ldots, (\bb k^{(N)}, W_N)$ is a large sequence of samples obtained from the weighted Feynman measure $\widetilde \mu_G$. We can also think of the algorithm performing a highly nontrivial variable transformation on Eq.~\eqref{eq:int1} that absorbs all propagator poles into the measure. This viewpoint is helpful when combining our algorithm with black-box integration methods (e.g. VEGAS \cite{Lepage:1977sw} or MadNIS \cite{Heimel:2022wyj}).

\subsection{Preprocessing}
\label{sec:pre}
As a \emph{subroutine}, the algorithm uses the tropical sampling algorithm from \cite{Borinsky:2020rqs} (compactly described in \cite[Algorithm 1]{Borinsky:2023jdv}). See~\ref{ap:example} for a detailed illustration of the algorithm via a worked-out example. For a gentle introduction to the tropical sampling procedure focusing on the application to Feynman integration, we also refer to \cite{andrea}. Before running this tropical sampling algorithm, a preprocessing step must be performed only once for a fixed graph, fixed edge weights, fixed spacetime dimension, fixed masses, and fixed momentum configuration. In this section, we summarize this initial step.

Each Feynman graph $G$ with $E$ edges has $2^E$ many subgraphs, as a subset of edges gives a subgraph. 
The following definition is due to Brown \cite[Def.~2.6]{Brown:2015fyf}. A subgraph $\gamma \subset G$ is called \emph{mass-momentum spanning} if the associated \emph{cograph} $G/\gamma$, which is the graph where all edges of $\gamma$ are contracted, is \emph{completely scaleless}. That means $G/\gamma$ has no mass dependence, and each vertex of $G/\gamma$ has zero total momentum flowing into it.
Equivalently, a subgraph $\gamma$ is mass-momentum spanning if and only if it contains all massive edges, and for each connected component of $\gamma$, the sum of all momenta flowing into the component is $0$. See \cite{Beekveldt:2020kzk} for a detailed discussion of the physical significance of mass-momentum-spanning subgraphs. Let $\delta_\gamma^{\mathrm{m.m.}}$ be $1$ if $\gamma$ is mass-momentum spanning and $0$ otherwise. Further, we write $L_\gamma$ for the number of loops of the subgraph $\gamma$.

In a preprocessing step, we compute tables $\omega$ and $J$ of size $2^E$ with one row for each subgraph $\gamma \subset G$, respectively. In functional notation, the $\omega$ table contains the data,
\begin{align} \label{eq:gen_dod} \omega(\gamma) = \sum_{e\in \gamma} \nu_e - D L_\gamma/2 - \omega_0 \cdot \delta_\gamma^{\mathrm{m.m.}} ~ \text{ for all } \emptyset \neq \gamma \subset G, \end{align}
with $\omega_0 = \sum_{e\in \gamma} \nu_e - D L_G/2$.
Defining the special case $\omega(\emptyset) = 1$ for the empty graph $\emptyset \subset G$ is convenient. The computationally most demanding step while filling this $\omega$ table is finding the number of loops $L_\gamma$ for each given subgraph $\gamma$.
The $J$ table is fixed recursively by $J(\emptyset)= 1$ and the rule
\begin{align} \label{eq:recursive} J(\gamma) = \sum_{e\in \gamma} \frac{J(\gamma\setminus e)}{\omega(\gamma\setminus e)} ~ \text{ for all } \emptyset\neq \gamma \subset G, \end{align}
where we sum over all edges of $\gamma$ and $\gamma\setminus e$ is the subgraph $\gamma$ with the edge $e$ deleted.

The terminal value of this recursion gives the prefactor $Z_G$ in Eq.~\eqref{eq:int2}-\eqref{eq:mc}:
\begin{align} Z_G = J(G) \cdot \frac{\Gamma(\omega_0)}{\prod_{e\in E} \Gamma(\nu_e)}, \end{align}
with the $\Gamma$-function and $\omega_0$ as defined above. The value $J(G)$ is an immediate generalization of the \emph{Hepp bound} associated to the graph $G$ \cite{Panzer:2019yxl}. Further, we assume that the values $L_\gamma$ and $\delta_\gamma^{\mathrm{m.m.}}$ are stored in memory for all subgraphs $\gamma\subset G$.

The total runtime necessary for the preprocessing is dominated by finding the number of loops for each subgraph $\gamma$. For graphs that are relevant in our context, computing the loop number takes linear time in the number of edges of $\gamma$, resulting in an overall runtime complexity of the order of $\mathcal{O}(E \cdot 2^E)$ for the preprocessing step. The memory complexity is obviously of the order of $\mathcal{O}(2^E)$, as all the tables have $2^E$ many rows of fixed size.

With these tables stored in memory (such that the contained data is accessible in constant time), we are ready to perform the following sampling algorithm.

\subsection{Sampling}
\label{sec:samp}

The following algorithmic steps produce a weight factor $W$ and a set of loop momenta $\bb k = (k_1,\ldots,k_L) \in (\R^D)^L$  distributed as the weighted Feynman measure $\widetilde \mu_G$ and ready to be used in formula~\eqref{eq:mc}.

\begin{enumerate}
\item \textbf{(Tropical sampling)}
The first step of the sampling algorithm is the generalized permutahedron tropical sampling algorithm \cite[Algorithm~4]{Borinsky:2020rqs} (see also Sec.~7.2 loc.~cit.). 

To run this algorithm, we need three variables $\gamma$, $\kappa$, and $\mathcal U^\tr$. The latter two contain numbers, the first variable $\gamma$ contains a subgraph (i.e., a subset of edges). We initialize $\gamma$ to contain all edges of $G$ and $\kappa = \mathcal U^\tr =1.$ As temporary storage variables, we also need one variable $x_e$ for each edge $e$ of $G$ and a variable $\mathcal V^\tr$. All these variables will be filled with numbers. The variables $\mathcal U^\tr$ and $\mathcal V^\tr$ are named consistently with a \emph{tropical geometric interpretation} of these variables (see \cite{Borinsky:2020rqs}). We do not need any knowledge of tropical geometry to run the algorithm. (However, such knowledge is essential while proving the algorithm's correctness and understanding the working principle.) Recall that we computed the values $J(\gamma)$, $\omega(\gamma)$, $L_\gamma$, and $\delta_\gamma^{\mathrm{m.m.}}$ for each subgraph $\gamma$ of $G$.

In the given order, we repeat the following steps until $\gamma$ contains no more edges:
\begin{enumerate}
\item 
Sample a random edge $e$ from the set $\gamma$ with probability $\frac{1}{J(\gamma)} \frac{J(\gamma\setminus e)}{\omega(\gamma \setminus e)}$.
\item Set $x_e = \kappa$ for the sampled edge $e$.
\item If $\delta_\gamma^{\mathrm{m.m.}}>\delta_{\gamma\setminus e}^{\mathrm{m.m.}}$, then set $\mathcal V^\tr =\kappa$.
\item If $L_\gamma > L_{\gamma\setminus e}$, then multiply $\mathcal U^\tr$ with $\kappa$ and store the result in $\mathcal U^\tr$, i.e.~$\mathcal U^\tr\leftarrow \mathcal U^\tr\cdot \kappa$.
\item 
Remove $e$ from the subgraph $\gamma$, i.e.~$\gamma \leftarrow \gamma \setminus e$.
\item Sample a uniformly distributed random number $\xi \in [0,1]$.
\item Multiply $\kappa$ with $\xi^{1/\omega(\gamma)}$ and store the result in $\kappa$, i.e.~$\kappa \leftarrow \kappa \cdot \xi^{1/\omega(\gamma)}$.
\item If $\gamma \neq \emptyset$, go back to step (a), otherwise return $x_1,\ldots,x_E$ and the values $\mathcal U^\tr$ and $\mathcal V^\tr$.
\end{enumerate}
The  sampling step (a) above is well-founded, because $ \sum_{e\in \gamma} \frac{1}{J(\gamma)} \frac{J(\gamma\setminus e)}{\omega(\gamma \setminus e)}=1 $
by Eq.~\eqref{eq:recursive}.
\item
Sample a random value $\lambda \in [0,\infty)$ following the  \emph{gamma distribution} with parameter $\omega_0$. The gamma distribution is given by the probability density
\begin{align} \frac{1}{\Gamma(\omega_0)} \lambda^{\omega_0} e^{-\lambda} \frac{\dd \lambda}{\lambda}. \end{align}
If $\omega_0$ is an integer $\geq 1$, then we can produce samples for $\lambda$ by drawing $\omega_0$ random numbers $\xi_1,\ldots,\xi_{\omega_0} \in [0,1]$ uniformly and setting $\lambda = - \sum_{k=1}^{\omega_{0}} \log \xi_k$.
However, more efficient and more general methods to sample from the distribution above exist (see, e.g., \cite[Sec.~3.4.1.E]{10.5555/270146}).
\item 
Sample $L$ vectors in $\R^D$ whose components are distributed normally. I.e.~sample $q_{1},\ldots,q_L \in \R^D$ using the measure
\begin{align} \prod_{\ell} \frac{\dd^D q_\ell }{(\sqrt{2\pi})^{D}} e^{ - \frac{q_\ell^2}{2} }, \end{align}
where $q_\ell^2 = (q_\ell^{(1)})^2 + \ldots + (q_\ell^{(D)})^2$ as usual. There are various standard methods to sample from the normal distribution (see, e.g., \cite[Sec.~3.4.1.C]{10.5555/270146}).
\item
Compute the $L\times L$ matrix $\mathcal L$ given by
\begin{align} \mathcal L_{\ell,\ell'} = \sum_e x_e \mathcal M_{e,\ell} \mathcal M_{e,\ell'}. \end{align}
\item
Compute the $L$ vectors $u_1, \ldots, u_L \in \R^D$ given by
\begin{align} u_\ell = \sum_e x_e \mathcal M_{e,\ell} p_e . \end{align}
\item Compute a Cholesky decomposition of $\mathcal L$: A lower triangular matrix $Q$ such that $\mathcal L = Q Q^T$.
\item 
Compute the values $\mathcal U, \mathcal V$ given by
\begin{gather} \mathcal U = \det \mathcal L = (\det Q)^2 \\  \mathcal V = \sum_e x_e (p_e^2 +m_e^2) - \sum_{\ell,\ell'} u_\ell \cdot u_{\ell'} \mathcal L^{-1}_{\ell,\ell'} .   \end{gather}
\item
Compute the weight factor
\begin{align} \label{eq:W} W = \left(\frac{\mathcal U^\tr}{\mathcal U}\right)^{D/2} \left( \frac{\mathcal V^\tr}{\mathcal V} \right)^{\omega_0}. \end{align}
\item
Compute $L$ vectors $k_1,\ldots, k_L$, given by
\begin{align} k_\ell = \sum_{\ell'} \left( \sqrt{\frac{\mathcal V}{2 \lambda} } (Q^T)^{-1}_{\ell, \ell'} q_{\ell'} - \mathcal L^{-1}_{\ell,\ell'} u_{\ell'} \right). \end{align}

\end{enumerate}

This finishes the algorithm. 
Following the above steps produces one set of momenta $\bb k = k_1,\ldots,k_L$ with the weight factor $W$ that can be used in  formula~\eqref{eq:mc}. That means, the pair $(\bb k,W)$ is distributed as the weighted Feynman measure $\widetilde \mu_G$.

\subsection{Bounds on the weight factor}
\label{sec:bounds}
The weight factor $W$ is a product of powers of 
quotients of polynomials with their tropical approximation.
These quotients are bounded by Theorem~3.3 of \cite{Borinsky:2020rqs}.
For some applications, it is convenient to have an explicit bound.

For a polynomial $p(x_1,\ldots,x_n) = \sum_{\bb i \in I} a_{\bb i} \prod_{j=1}^n x_j^{i_j}$ with coefficients $a_{\bb i}\neq 0$ that are multi-indexed by $\bb i = (i_1,\ldots,i_n)$ within some given finite index set $I$, we define the \emph{tropical approximation} of $p$ by setting $p^\tr(x_1,\ldots,x_n) = \max_{i \in I} \prod_{j=1}^n x_j^{i_j}$. 
If all the coefficients $a_{\bb i}$ are positive, then we have $a_{\mathrm{min}} \cdot p^\tr(x_1,\ldots,x_n) \leq p(x_1,\ldots,x_n)$ for all $x_j \geq 0$ with $a_{\mathrm{min}}$ being the minimal coefficient of the polynomial $p$. Moreover, $p(x_1,\ldots,x_n) \leq \left( \sum_{\bb i\in I} a_{\bb i} \right) \cdot p^\tr(x_1,\ldots,x_n)$ for all $x_j \geq 0$. Under certain conditions, similar inequalities hold in the case where some coefficients $a_{\bb i}$ are negative (see \cite[Theorem~3.3]{Borinsky:2020rqs}).

The values $\mathcal U, \mathcal U^\tr, \mathcal V$, and $\mathcal V^\tr$ that are used in the algorithm above are closely related to the \emph{Symanzik polynomials} in Feynman parameter space and the associated tropical approximation. See, e.g., \cite{Borinsky:2023jdv} for a more detailed account of the Symanzik polynomials.
The $\mathcal U$ polynomial is defined by $\mathcal U(x_1,\ldots,x_E) = \sum_{T \subset G} \prod_{e \not \in T} x_e$, where we sum over all spanning trees of the Feynman graph $G$. The $\mathcal F$ polynomial is given by 
$\mathcal F(x_1,\ldots,x_E) = \sum_{F \in G} p(F)^2 \prod_{e \not \in F} x_e + \mathcal U \cdot \sum_{e} x_e m_e^2$, where we sum over all spanning 2-forests $F$ of the graph and $p(F)$ is the momentum flowing between the two components of the forest.

All coefficients of the $\mathcal U$ polynomial are $1$.
Hence, we always have ${1}/{\# T}\leq\mathcal U^\tr/ \mathcal U\leq 1$,
where $\# T$ is the number of spanning trees of the graph.
Further let $P_{\textrm{min}}^2$ be the minimal coefficient of the 
$\mathcal F$ polynomial. We can also think of $P_{\textrm{min}}^2$ 
as the \emph{minimal scale} of the Feynman graph.
Hence, as all coefficients of $\mathcal F$ are positive because we are in an effectively Euclidean regime, $P_{\textrm{min}}^2 \cdot \mathcal F^\tr \leq \mathcal F$ for all
$x_e \geq 0$.

The value of $\mathcal V$ is defined as the quotient $\mathcal V=\mathcal F/\mathcal U$ and $\mathcal V^\tr = \mathcal F^\tr/\mathcal U^\tr$ (see, e.g.,~\cite{Borinsky:2023jdv} for details). Recall that we require $\omega_0 > 0$ for convergence. Hence, by combining the previous inequalities with Eq.~\eqref{eq:W}, we get 
\begin{align} W \leq W_{\mathrm{max}}= \begin{cases} P_{\textrm{min}}^{-2\omega_0} & \text{ if } D/2 \geq \omega_0 \\
(\# T)^{\omega_0-D/2} P_{\textrm{min}}^{-2\omega_0} &\text{ else} \end{cases} \end{align}
The second case, relevant if $\omega_0 > D/2$, is quite pessimistic for larger graphs. We expect that it can be replaced with a more efficient bound.

\subsection{Sampling directly from the Feynman measure}

With the bounds on the weight factor $W$, we can produce samples from the unweighted Feynman measure $\mu_G$~\eqref{eq:mu}. The method is a standard \emph{acceptance/rejection sampling} approach:

\begin{enumerate}
\item Sample a pair of loop momenta and weight factor $\bb k, W$ distributed as the density $\widetilde \mu_G$ via the algorithm described in Sections~\ref{sec:pre}--\ref{sec:samp}.
\item Draw a uniformly distributed random number from the interval $\xi \in [0,W_{\mathrm{max}}]$.
\item Return $\bb k$ if $\xi < W$, otherwise reject the sample and start again at 1.
\end{enumerate}

These steps produce samples distributed as the Feynman measure $\mu_G$ in Eq.~\eqref{eq:mu}. 
It depends on the concrete integration problem if it is more efficient to use the weighted Feynman measure $\widetilde \mu_G$ directly as explained in Section~\ref{sec:overview} or the Feynman measure $\mu_G$ in Eq.~\eqref{eq:mu}.
If evaluating the integration kernel $f(\bold{k})$ is computationally demanding, the extra steps might be worth producing samples from the unweighted measure $\mu_G$.
If $f(\bold{k})$ can be quickly evaluated, then using $\widetilde \mu_G$ is likely more efficient. For the example problems we discuss below, the evaluation of $f(\bold{k})$ is relatively cheap, so we use  the weighted measure $\widetilde \mu_G$ and skip the acceptance/rejection step above.

\subsection{Proof of the algorithm}

To explain the inner workings and prove the correctness of the algorithm, we start with Eq.~\eqref{eq:int1} for the integral $I_{G,f}$ that depends on the graph $G$ with all the external momentum data, masses and edge weights and the function $f$ of the loop momenta. Applying the Schwinger parameterization 
\begin{align} \frac{1}{D_e^{\nu_e}} = \frac{1}{\Gamma(\nu_e)} \int_0^\infty \frac{\dd x_e'}{x_e'} {x_e'}^{\nu_e} e^{-x_e' D_e} \end{align}
to every propagator in Eq.~\eqref{eq:int1} results in
\begin{align} I_{G,f} = \left(\prod_{e=1}^{E} \int_{0}^{\infty} \frac{\dd{x_e'}}{x_e'} \frac{{x_e'}^{\nu_e } }{\Gamma(\nu_e)} \right) \int_{\R^{DL}} \prod_{\ell = 1}^{L} \frac{\dd^D{{k}_{\ell}}}{\pi^{D / 2}} \exp\left(-\sum_{e=1}^E x_e' D_e\right) \cdot f({\bb k}) \; \end{align}
To go from Schwinger parameters to Feynman parameters, we use the identity $1 = \int_0^{\infty} \dd{t} \delta(t - \sum_e x'_e)$ and apply the coordinate transformation $x'_e = t x_e$. The effective domain of
integration over $x_e$ is now the unit simplex instead of $\mathbb{R}_{+}^E$. 
\begin{align} I_{G,f} = \int_0^\infty \frac{\dd{t}}{t} \bigg( \prod_{e} \int_{0}^{\infty} \frac{\dd x_e}{x_e} \frac{x_e^{\nu_e} }{\Gamma(\nu_e)} \bigg) t^{\sum_e \nu_e} \delta(1 - \sum_e x_e) \\ \cross \int_{\R^{DL}} \bigg( \prod_{\ell} \frac{\dd^D{{k}_{\ell}}}{\pi^{D / 2}} \bigg) \exp(- t \sum_{e} x_e D_e) \cdot f(\bb k) , \end{align}
where we sum or multiply over all edges $e$ or loops $\ell$ of $G$.

We now rewrite the argument of the exponential in terms of vectors and matrices. Recall the definition of $D_e$ in Eq.~\eqref{eq:prop}. The quadratic part of the argument of the exponential is governed by an $L \cross L$ matrix $\mathcal{L}$ whose components
are given by 
\begin{equation}
\label{eq:l_matrix}
    \mathcal{L}_{\ell, \ell'} = \sum_e x_e \mathcal{M}_{e,\ell} \mathcal{M}_{e,\ell'} \; ,
\end{equation}
This matrix is related to the first Symanzik polynomial $\mathcal{U}$
via the determinant:
\begin{equation}
\label{eq:u_polynomial} 
    \mathcal{U} = \det \mathcal{L} \; .
\end{equation}
The linear part of $- t \sum_{e} x_e D_e$ is determined by a set of $L$ vectors $u_\ell$ defined as 
\begin{equation}
\label{eq:u_vec}
    u_\ell = \sum_e x_e \mathcal{M}_{e, \ell} p_e \;.
\end{equation}
With these definitions, we can express the argument as follows:
\begin{equation}
\label{eq:exp_quad}
    - t \sum_e x_e D_e = - t \left( \bb k^T \mathcal{L} \bb k + 2 \bb u^T \bb k + \sum_e {x_e} (p_e^2 + m_e^2) \right) \; ,
\end{equation}
where the bold letters stand for vectors of vectors whose entries are multiplied via the Euclidean scalar product. Completing the square of this quadratic allows us to find the coordinate transformation required to simplify the exponential to a normal Gaussian distribution. To complete the square, we introduce the following rational function in $\bb x$:
\begin{equation}
\label{eq:v_polynomial} 
\mathcal{V} = \sum_e {x_e} (p_e^2 + m_e^2) 
- \bb u^T \mathcal{L}^{-1} \bb u \; .
\end{equation}
$\mathcal{V}$ is related to both the first Symanzik polynomial $\mathcal U$ and the second Symanzik polynomial $\mathcal{F}$ through the relation $\mathcal{V} = \frac{\mathcal{F}}{\mathcal{U}}$, as explained in the last section. Inserting $\mathcal{V}$ into Eq.~\eqref{eq:exp_quad} gives
\begin{equation}
    - t \sum_e x_e D_e = - t \left( \bb k^T \mathcal{L} \bb k + 2 \bb u^T \bb k + \bb u^T \mathcal{L}^{-1} \bb u \right) - t \mathcal{V} \; .
\end{equation}
The next step is another change of variables: $t = \frac{\lambda}{\mathcal{V}}$. After
applying this coordinate transformation $-t \sum_e x_e D_e$ now reads

\begin{equation*}
    - t \sum_e x_e D_e = - \frac{\lambda}{\mathcal{V}} \left( \bb k^T \mathcal{L} \bb k + 2 \bb u^T \bb k + \bb u^T \mathcal{L}^{-1} 
    \bb u \right) - \lambda \; .
\end{equation*} 
The Jacobian of this transformation moves $\mathcal{V}$ from the exponential to the 
denominator:

\begin{equation*}
    \dd{t} t^{\sum_e \nu_e - 1} = 
    \dd{\lambda} \lambda^{\sum_e \nu_e - 1} \frac{1}{\mathcal{V}^{\sum_e \nu_e}} \;  .
\end{equation*}
We can now complete the square using new coordinates $\bb q\in \R^{DL}$
\begin{equation}
\label{eq:loop_momenta}
    \bb k = \sqrt{\frac{\mathcal{V}}{2\lambda}} (Q^T)^{-1} \bb q - \mathcal{L}^{-1} \bb u  \;,
\end{equation}
where the matrix $Q$ is a factorization of $\mathcal L$, for instance obtained using the Cholesky decomposition, such that $\mathcal{L} = Q^T Q$ and $\det Q = \mathcal{U}^{1/2}$. Hence, we have the following relation between the measures in $\bb k$ and the measure in $\bb q$: 
\begin{equation*}
    \prod_\ell \frac{d^D k_\ell}{\pi^{D/2}} = \left(\frac{\mathcal{V}}{\lambda}\right)^{DL / 2} 
    \frac{1}{\mathcal{U}^{D/2}}
    \prod_\ell \frac{d^D q_\ell}{(2\pi)^{D/2}} \;.
\end{equation*}
Combining all these transformations gives the following representation of the integral \eqref{eq:int1} in which the Gaussian nature of the loop momenta becomes explicit:
\begin{equation}
\begin{split}
   I_{G,f} = \int_0^\infty \frac{\dd{\lambda}}{\lambda}  \lambda^{\omega_0} e^{-\lambda} \int_{x_e \geq 0} \left(\prod_{e}^{E} \frac{x^{\nu_e-1} \dd x_e}{\Gamma(\nu_e)} \right) 
    \frac{\delta(1-\sum_e x_e)}{\mathcal{U}^{D/2} \mathcal{V}^{\omega_0}} 
 \times \\\int \prod_{\ell}
 \frac{\dd^D{{q}_{\ell}}}{(2 \pi)^{D / 2}} \exp\left(- \sum_{\ell }\frac{q_\ell^2}{2}\right) f(\bb k(\bb q, x_e, \lambda)) \; .
\end{split}
\end{equation}
Notice that in the integral representation above, we removed the integrable singularities originating from the propagators but introduced new singularities from the $\mathcal U$ and $\mathcal V$ terms.
We will deal with these remaining singularities using the tropical sampling approach.

In the case where $f(\bb k)=1$, we would now be able to analytically perform the integrals over $\lambda$ and $\bb q$, but here the complicated structure of $f(\bb k)$ prevents this step. Instead, we can identify several probability distributions in this expression. These probability distributions come with known algorithms to sample from them. We can use these algorithms to numerically evaluate the integral $I_{G,f}$.

First, we may recognize the gamma distribution, which has the probability density:
\begin{equation}
    \label{eq:gamma_dist}
    \frac{1}{\Gamma(\omega)} \lambda^{\omega} e^{-\lambda} \frac{\dd \lambda}{\lambda}\; .    
\end{equation}
The second term can be interpreted as a perturbed version of the tropical measure $\mu^\tr$ from \cite{Borinsky:2020rqs}:
\begin{equation*}
\begin{split}
\left(\prod_{e} x^{\nu_e-1} \dd x_e \right) 
    \frac{\delta(1-\sum_e x_e)}{\mathcal{U}^{D/2} \mathcal{V}^{\omega_0}} 
=
 J(G) 
    \left(\frac{\mathcal{U}^{\tr}}{\mathcal{U}}\right)^{D/2} \left(\frac{\mathcal{V}^{\tr}}{\mathcal{V}
    }\right)^{\omega_0} \mu^{\tr},
\end{split}
\end{equation*}
where we sample over the positive simplex in the $x_e$ coordinates, or equivalently, over $E-1$ dimensional positive projective space. Algorithm 4 in \cite{Borinsky:2020rqs} produces samples from the measure $\mu^\tr$. The correction factor that consists of quotients of Symanzik polynomials and their tropical approximations on the right-hand side can be identified with the weight factor $W$. The normalization factor $J(G)$ is computed as explained in Section~\ref{sec:pre}. Its value is also called $I^\tr$ in~\cite{Borinsky:2020rqs}.

We can also recognize the $DL$-dimensional normal distribution:
\begin{equation*}
    \prod_{\ell}  \exp\left( -\frac{q_\ell^2}{2}\right) \frac{\dd q_\ell}{(2\pi)^{D/2}} \;.
\end{equation*}
Except for the tropical sampling measure, all these probability distributions are elementary. Sampling from them and making the necessary substitutions to recover $\bb k$ and $W$, we recover the algorithm described in Sections~\ref{sec:pre}--\ref{sec:samp}.

\section{Implementation}
\label{sec:imp}

\subsection{The \texttt{momtrop} package}

We implemented the algorithm from Sections \ref{sec:pre} and \ref{sec:samp} in the \verb|Rust| programming language as a standalone library named \verb|momtrop| (\url{https://github.com/alphal00p/momtrop}). 
The \verb|momtrop| library can be used directly and conveniently as a sampling method within the $\gamma$Loop package (\url{https://github.com/alphal00p/gammaloop}).
$\gamma$Loop is the successor of $\alpha$Loop~\cite{Capatti:2020xjc}, and aims to automate the computation of differential cross-sections and IR-finite amplitudes. 

In order to use \verb|momtrop|, 
first install \verb|Rust| (\url{https://www.rust-lang.org/tools/install}) and \verb|git| (\url{https://git-scm.com/}). Then clone the \verb|momtrop| repository by running the command
\begin{verbatim}
    git clone git@github.com:alphal00p/momtrop.git
\end{verbatim}
After switching to the \verb|momtrop| directory, The command
\begin{verbatim}
    cargo test triangle --release -- --nocapture
\end{verbatim}
installs all dependencies, runs the triangle example from Section~\ref{sec:triangle} with $N=10^6$ sample points and prints the result. The source code for this example can be found in \verb|/tests/triangle.rs|. 

The library is designed with maximum flexibility in mind, and thus defers the actual evaluation of the integrand to the user. 
The \verb|momtrop| package can be used in any \verb|Rust| project by running \verb|cargo add momtrop| in your project directory. 
\verb|momtrop| takes a Feynman graph topology with kinematic data as input. For each topology, \verb|momtrop| produces a sampler (i.e.,~a sampling routine) that outputs tuples of vectors $\bb k =(k_1,\ldots,k_L)$ together with the weight factor $W$. The user can then use these samples to evaluate integrals such as \eqref{eq:int1}. 

The topology is provided as a list of edges, which are sets of vertices with a weight $\nu_e$ and a boolean which tells \verb|momtrop| if the edge is massive. For example, to create a massive edge connecting  the vertices 0 and 1 with weight $\nu=2/3$, we need the code
\begin{verbatim}
    let edge = Edge {
        vertices: (0, 1),
        is_massive: true,
        weight: 2. / 3.,
    };
\end{verbatim}
As usual, each edge must have some loop momentum flowing through it. We also need to specify which vertices have a leg attached. These vertices are provided as a list. For example, to create the massless triangle topology, we use the following code: 
\begin{verbatim}
    let weight = 5. / 6.;

    let triangle_edges = vec![
        Edge {
            vertices: (0, 1),
            is_massive: false,
            weight,
        },
        Edge {
            vertices: (1, 2),
            is_massive: false,
            weight,
        },
        Edge {
            vertices: (2, 0),
            is_massive: false,
            weight,
        },
    ];

    let externals = vec![0, 1, 2];

    let graph = Graph {
        edges: triangle_edges,
        externals,
    };
\end{verbatim}
Any information about edges that is provided at a later stage is 
expected to be ordered in the same manner as the list that is provided to 
the \verb|Graph| struct as above. 

We must specify how the loop momenta are routed through the graph to build the sampler. We do so by providing the loop-incidence matrix $\mathcal{M}_{e,\ell}$ as defined in Eq.~\eqref{eq:prop}. The following code builds the sampler for the triangle example: 
\begin{verbatim}
    let loop_signature = vec![vec![1]; 3];
    let sampler = graph.build_sampler(loop_signature)?;
\end{verbatim}
The function \verb|build_sampler| returns a \verb|Result| type.  Building the sampler might be impossible due to a subdivergence. In this case, \verb|build_sampler| returns the \verb|Err| variant. 

If \verb|build_sampler| is successful, the resulting \verb|SampleGenerator| object can be used to generate sample points. For maximum flexibility, the user can provide their own set of uniform random numbers in the unit interval to the sampler. The method \verb|.get_dimension()| can be called on a \verb|SampleGenerator| object to determine how many uniform random numbers are needed. 

The number of uniform random variables is counted as follows: For a given graph $G$ with $E$ edges and $L$ loops, sampling from $\mu^{\tr}$ with spacetime dimension $D$ requires $2E - 2$ uniform random variables. The gamma distribution requires just a single uniform random variable. The standard normal distributions require $DL + (DL \textrm{ mod } 2)$ uniform random variables. Adding $DL \textrm{ mod } 2$ is necessary since we internally use the Box-Müller method to produce normally distributed samples in pairs. In total, the number of uniform random variables the algorithm requires for each sample point is thus $2E - 1 + DL + (DL \textrm{ mod } 2)$.

To produce a sample point, we also need to provide the mass of each edge along with the external-dependent shift $\vec{p}_e$ as defined by Eq.~\eqref{eq:prop}. The mass is provided as an option type where the \verb|None| variant should be used for massless edges. The external shifts are provided by a \verb|Vector| type specific to \verb|momtrop|. A \verb|Vector| can be easily constructed from \verb|Rust|'s \verb|Array| type. For example, to create the edge data for the massless triangle topology, the following code is required:
\begin{verbatim}
    let p1 = Vector::from_array([3.0, 4.0, 5.0]);
    let p2 = Vector::from_array([6.0, 7.0, 8.0]);

    let edge_data = vec![
        (None, Vector::from_array([0.0, 0.0, 0.0])),
        (None, p1),
        (None, (&p1 + &p2)),
    ];
\end{verbatim}
Generating a point also requires a settings struct. To generate the 
default settings, simply use  \verb|TropicalSamplingSettings::default()|. 

A point is generated by calling the method
\begin{verbatim}
     generate_sample_from_x_space_point(x_space_point, edge_data, settings)
\end{verbatim}
of the \verb|SampleGenerator| object. The variable \verb|x_space_point| contains the list of input uniform random variables.
\verb|generate_sample_from_x_space_point| returns the \verb|Result| type. If problems are encountered in the algorithm, the method returns the \verb|Err| variant. This happens, for example, when the Laplacian $\mathcal{L}$ has a vanishing determinant. If there is no problem, a \verb|TropicalSamplingResult| object is returned, which contains the generated loop momenta. It also contains the field \verb|jacobian|, which is equal 
to $Z_G \cdot W \cdot \pi^{DL/2}$. With this data, one can use~\eqref{eq:mc} to estimate the integral. The 
factor of $\pi^{DL/2}$ is included such that it is easier to switch to a different convention.

The dimension $D$ is implemented using \verb|Rust|'s constant generics. In most examples, the compiler can automatically infer this parameter from the vectors constructed by the user. Moreover, the code is generic over floating point types such that the algorithm can be run in quadruple or arbitrary precision. 

For a fully worked-out example, see the file \verb|tests/triangle.rs| on the GitHub page. Further documentation on the \verb|momtrop| package can be found on \url{https://docs.rs/momtrop/latest/momtrop/}.

\subsection{Performance discussion}
What follows is a naive method to map an Euclidean loop integral in $D=3$ to the unit hypercube. Our algorithm enables integration with a much lower variance than with this naive technique. In this section, we compare the runtime of our sampling algorithm with this basic method.
We can parametrize each loop momentum $k_\ell$ in the following manner:
\begin{equation}
\label{eq:spherical_param}
    k_\ell(x_\ell, y_\ell, z_\ell) = b \cdot Q \left (\frac{1}{1-x_\ell} - \frac{1}{x_\ell} \right) \begin{pmatrix} 
    2 \sqrt{z_\ell (z_\ell - 1)} \cos(2\pi y_\ell) \\ 
    2 \sqrt{z_\ell (z_\ell - 1)} \sin(2\pi y_\ell) \\ 
    -1 + 2 z_\ell \\
    \end{pmatrix} \; , 
\end{equation}
where $(x_\ell, y_\ell, z_\ell) \in [0,1]^3$, $Q$ is a typical energy scale, and $b$ is a tunable dimensionless parameter. This parameterization is related to the spherical coordinate system 
$(r, \phi, \theta)$ in the following way:
\begin{equation*}
\begin{split}
    r_{\ell} &= b \cdot Q \left (\frac{1}{1-x_\ell} - \frac{1}{x_\ell} \right) \;, \\
    \phi_\ell &= 2 \pi y_\ell  \,, \\
    \cos(\theta_\ell) &= -1 + 2 z_\ell \,.
\end{split}
\end{equation*}
Disregarding sets of measure $0$, this coordinate transformation maps $\R^3$ to the unit cube bijectively. The advantage of mappings as the one above is their simplicity. 
Measured on an Intel Xeon W-2135 CPU, it takes roughly $ 400\unit {ns}$ to perform the variable transformation~\eqref{eq:spherical_param} for a full-fledged Feynman integral. The major disadvantage of transformation rules, as~\eqref{eq:spherical_param}, is that they create new, spurious integrable singularities of the integrand. It is hard to tame these singularities numerically. 
The algorithm from Sections \ref{sec:pre} and \ref{sec:samp} that is implemented within \texttt{momtrop} avoids the introduction of new spurious singularities completely and is therefore much better behaved numerically.
We emphasize again that previous approaches, even though they rely on the naive integration method, are typically enhanced by black-box variance reduction techniques such as VEGAS~\cite{Lepage:1977sw} or multi-channeling~\cite{Becker:2012aqa}. Here, we compare the new tropical sampling approach only to the naive baseline without any variance reduction, so the comparison should not be interpreted as wholly representative of the state of the art. Our experiments suggest that the tropical sampling method also performs very favorably compared to enhanced applications of the naive method; see, e.g., Section~\ref{sec:multi} where we compare tropical sampling with a multi-channelling enhanced approach. We also remark that black-box variance reduction methods could also be combined with the tropical sampling approach. While increasing the time per sample, this will also reduce the variance even further. We will not attempt to combine tropical sampling with further variance reduction methods in this article.

The time it takes to generate a sample point with \texttt{momtrop} depends heavily on the complexity of the topology. The timings range from around $1\unit{\us}$ for the simplest example to around $15 \unit{\us}$ for the most complicated example. This is significantly slower than the naive parameterizations of the integral in \eqref{eq:spherical_param}.
However, the negative performance impact is expected to be minimal for physical applications, where the evaluation time of $f(\mathbf{k})$ dominates by several orders of magnitude. See Section~7.4 of \cite{Capatti:2022tit} for examples of such timings.
Further, the tropical sampling approach reduces the variance of the Monte Carlo integration process drastically. Thereby, significantly fewer samples are necessary. This effect typically overcompensates the longer runtime of the tropical sampling algorithm by multiple orders of magnitude (see the benchmarks in Section~\ref{sec:ltd}).

Within \texttt{momtrop}, samples from the gamma distribution are generated using the so-called inverse CDF method, using a custom implementation of an algorithm for the evaluation of the inverse lower incomplete gamma function described in \cite{incomplete_gamma}. The runtime of this algorithm depends on the parameter $\omega$. The timing is usually between $300 \unit{ns}$ and $1 \unit{\mu s}$. In the case $\omega =1$, the lower incomplete gamma function degenerates to a logarithm, giving a sub-nanosecond timing. Although there are more efficient methods for sampling from the gamma distribution, this method has the advantage that a \emph{fixed} number of uniform random variables is computed per sample point. This ensures we can add adaptive sampling methods on top of the procedure described in this paper at a later point.

\section{Visualization of the measures and importance sampling}
\label{sec:visual}
To illustrate the efficacy of the algorithm and the implementation, we visualize the weighted Feynman probability measure $\widetilde \mu_G$ as defined by Eq.~\eqref{eq:int2}. To create these visualizations for a specific topology, we generate many samples with \texttt{momtrop} and create a histogram by binning the generated loop momenta into a two-dimensional grid. These histograms are then displayed as heatmaps. This way, we visualize how our algorithm favours samples near the integrable singularities of the Feynman integral. We only show the result of the momentum sampling and neglect the weight factor $W$ in these visualizations. The code used to create these visualizations can be found on (\url{https://github.com/alphal00p/tropical-plots}).

\begin{figure}[ht]
    \centering
    \includegraphics[width=0.8\linewidth]{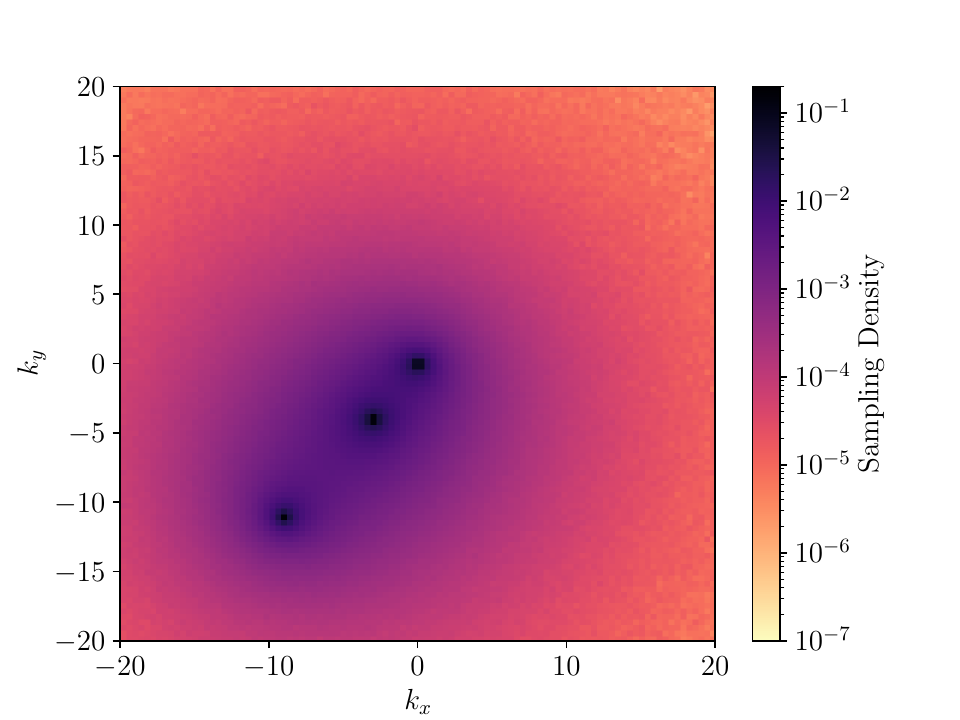}
    \caption{$\widetilde \mu_G$ for the triangle topology in $D=2$. 
    }
    \label{fig:triangle_plot}
\end{figure}

In Figure~\ref{fig:triangle_plot}, we visualize $\widetilde \mu_G$ for the triangle topology, depicted in \eqref{eq:triangle}, in $D=2$ spatial dimensions with the kinematics %
\begin{equation*}
\begin{split}
    \vec{p}_1 &= (3, 4), \\
    \vec{p}_2 &= (-6, -7), \\ 
    \vec{p}_3 &= \vec{p}_1 - \vec{p}_2,
\end{split}
\end{equation*} 
edge weights $\nu_e=2/3$, and the loop momentum routing as in \eqref{eq:triangle}.
For those kinematics, the integrable singularities are located at $\vec{k}=0$, $\vec{k} = (-3, -4)$ and $\vec{k} = (-9, -11)$.  The heatmap shows the sampling density 
at each specific $\vec{k}=(k_x,k_y)$ coordinate. It can be seen that the sampling density increases significantly close to the integrable singularities.

We also visualize the double triangle topology
\begin{align} \begin{tikzpicture} \begin{feynman} \vertex(int1); \vertex[above right=2cm and 2cm of int1](int2); \vertex[below right=2cm and 2cm of int1](int3); \vertex[below right=2cm and 2cm of int2](int4); \vertex[left=of int1](ext1); \vertex[right=of int4](ext2); \diagram*[large]{ (int2) -- [--,line width=0.4mm, black, momentum=\(k\)] (int1), (int1) -- [--,line width=0.4mm, black, edge label=\(m\)] (int3), (int2) -- [--,line width=0.4mm, black] (int3), (int4) -- [--,line width=0.4mm, black, momentum=\(l\)] (int2), (int3) -- [--,line width=0.4mm, black, edge label=\(m\)] (int4), (ext1) -- [--,line width=0.4mm, black, momentum=\(p_1\)] (int1), (int4) -- [--,line width=0.4mm, black, momentum=\(p_1\)] (ext2), }; \end{feynman} \end{tikzpicture} \end{align}
in $D=1$ with the loop momentum routing and two massive edges as depicted. Due to the two massive edges, only three out of five propagators may vanish. We do not need to specify the masses, as their precise value does not influence the sampling algorithm. The external momentum is
$ \vec{p}_1 = (1)$
and the edge weights are set to $\nu_e=3/10$.

\begin{figure}[ht]
    \centering
    \includegraphics[width=0.8\linewidth]{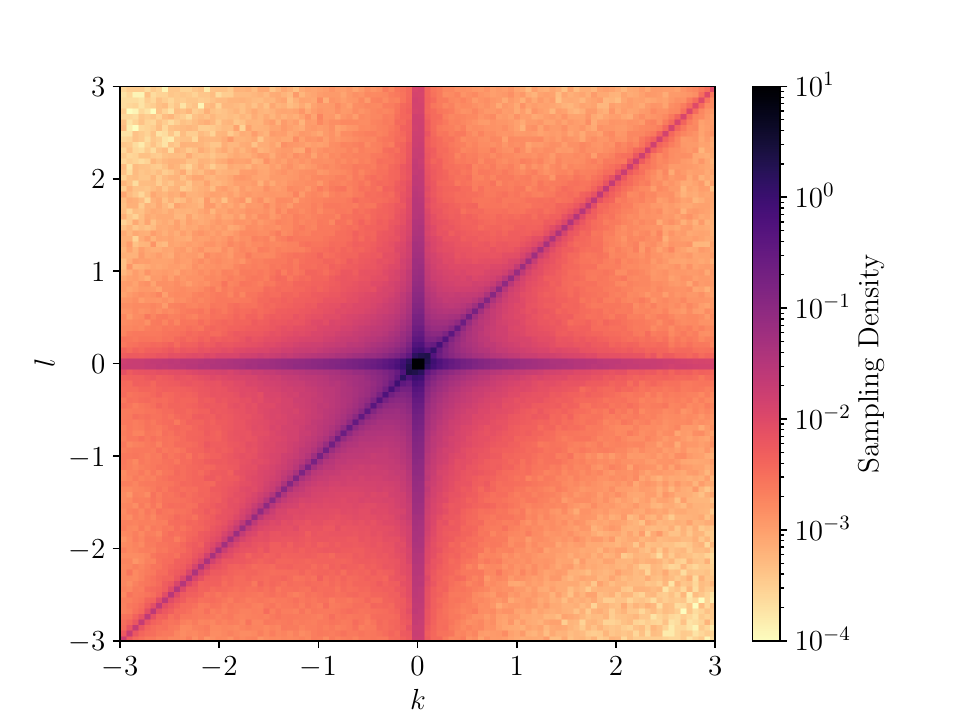}
    \caption{$\widetilde \mu_G$ for the double triangle topology in $D=1$.} 
    \label{fig:double_triangle_plot} 
\end{figure}

Figure~\ref{fig:double_triangle_plot} shows the resulting heatmap.
Here, the horizontal axis shows the loop momentum $k$ and the vertical axis the momentum $l$. For the provided kinematics, the integrable singularities are located along three lines defined by $k = 0$, $l = 0$ and $k = l$. The sampling density along these lines is clearly enhanced. %

\section{Application to loop-tree duality and examples}
\label{sec:ltd}
Integrals of the form~\eqref{eq:int1} appear in \emph{loop-tree duality} (LTD)~\cite{Catani:2008xa, Sborlini:2016gbr,  Sborlini:2016hat, Capatti:2019ypt, Capatti:2019edf, deLejarza:2024scm}, which is a numerical approach to computing loop integrals and differential cross-sections~\cite{Soper:1999xk, Gong:2008ww, Capatti:2020xjc, Capatti:2022tit}. LTD was successfully applied to light-by-light scattering \cite{AH:2023kor}, QCD at finite chemical potential \cite{Navarrete:2024zgz}, and vector boson production \cite{Kermanschah:2024utt}. 

The Feynman integral in expression~\eqref{eq:feyn} with $D=4$ and $\nu_e=1$ serves as the starting point of LTD. The 4-dimensional loop integral is then reduced to an integral over 3-dimensional loop momenta by analytically integrating each energy component $k^0_{\ell}$. The remaining integral over the spatial loop momenta $k _{\ell}$ is performed with Monte Carlo methods. There exists a multitude of algorithms \cite{Capatti:2019ypt, Capatti:2020ytd, Capatti:2022mly} that perform the analytic integration over $k^0_{\ell}$ for an arbitrary graph $G$. These algorithms provide different representations of the same integrand. All representations are identical as functions of the spatial loop momenta, but they reveal different properties of the LTD integrand and have varying numerical properties. 

A particularly convenient representation is the cross-free-family (CFF) representation~\cite{Capatti:2022mly}, which is free from spurious singularities. The CFF integral for a graph $G$ naturally takes the form of equation~\eqref{eq:int1} with $D=3$ and $\nu_e = 1/2$. If we restrict ourselves to finite topologies in the Euclidean regime, the associated function $f(\bold{k})$ is well-behaved.  

Note that by including a factor of $\prod_{e=1}^E D_e(\bb k)^{\mu_e}$ into the integration kernel $f(\bb k)$ in Eq.~\eqref{eq:int1}, we can increase or decrease the propagator powers $\nu_e$ in the same equation. Such manipulations allow us to tackle a broader range of integration problems. Even though the result of the integration does not depend on how the integrand is split between kernel and measure, the variance of the numerical integration very much depends on this splitting. In the following examples, we use this freedom to improve convergence rates, but we leave the problem of finding the \emph{optimal} split for future work. 

Below, we give a series of examples that test the algorithm in the context of LTD. $I$ denotes the value of the integral defined by  Eq.~\eqref{eq:feyn} in Minkowski space and $D=4$. 
An application of LTD transforms this into  Eq.~\eqref{eq:int1}, now with a Euclidean metric and $D=3$. Unless stated otherwise, we use $N=10^9$ sample points. $I_{\text{ref}}$ denotes a reference value of the integral under inspection obtained with an alternative method. The deviation from this reference value is denoted by $\Delta$ and is reported in terms of the statistical error $\sigma$ and as a percentage of the central value. The time spent in the tropical sampling algorithm $T_{\text{tr}}$ and the time spent in the evaluation of the LTD expression $T_{\text{ltd}}$ for a single sample point are also reported. All timings are measured on an Intel Xeon W-2135 CPU. 

To demonstrate the advantages of our new tropical approach, we compare it to the result of the integration when the naive parameterization~\eqref{eq:spherical_param} is used. We denote the naive integration result by $I_n$. The error and deviation are listed as $\sigma_n$ and $\Delta_n$. Further, we report the squared relative variance $C^2 = \frac{\mathrm{Var}[I]}{I_{\text{ref}}^2} = \frac{N \sigma^2}{I_{\text{ref}}^2}$, $C_n^2 = \frac{N \sigma_n^2}{I_{\text{ref}}^2}$  which give a dimensionless metric for the convergence speed. For example, if we find that $C_n^2 \approx 2 C^2$, we can infer that the naive method requires about twice as many sample points to compute $I$ to the same accuracy as \texttt{momtrop}. 

Besides parameterization~\eqref{eq:spherical_param}, other bijective maps from $\mathbb{R}^3$ to the unit cube may be used instead. Most older methods, including $\gamma$loop, support a variety of such mappings. For some of the examples we present below, a different map would surely be better suited, resulting in a faster convergence for the naive method. More importantly, most older methods combine such maps with adaptive sampling. In the following examples, we compare our tropical sampling method to the entirely naive approach without adaptive sampling. Our justification for this is that, firstly, the proper choice of an adaptive sampling strategy is more of an art than an engineering problem. There is a huge number of methods, each of which comes with free parameters that need to be optimized externally. Our tropical sampling approach has the advantage of being dedicated to the Feynman measure problem, and it comes with a relatively small number of free parameters that need to be chosen. Secondly, we could also put adaptive sampling on top of our tropical sampling method, which would likely still increase the efficacy of our approach even further.

\subsection{The triangle example: A 1-loop 3-point topology}
\label{sec:triangle}

We start with the massless triangle topology: %
\begin{align} \begin{tikzpicture} \begin{feynman} \vertex(int1); \vertex[above right=2cm and 2cm of int1](int2); \vertex[below right=2cm and 2cm of int1](int3); \vertex[left=of int1](ext1); \vertex[right=of int2](ext2); \vertex[right=of int3](ext3); \diagram*[large]{ (int1) -- [--,line width=0.4mm, black] (int2), (int1) -- [--,line width=0.4mm, black] (int3), (int2) -- [--,line width=0.4mm, black] (int3), (ext1) -- [--,line width=0.4mm, black, momentum=\(p_1\)] (int1), (int2) -- [--,line width=0.4mm, black, momentum=\(p_2\)] (ext2), (int3) -- [--,line width=0.4mm, black, momentum=\(p_3\)] (ext3), }; \end{feynman} \end{tikzpicture} \label{eq:triangle} \end{align}
The external momenta are set to: 
\begin{equation*}
\begin{split}
    p_1 &= (1.0, 3.0, 4.0, 5.0) \;, \\
    p_2 &= (-1.0, -6.0, -7.0, -8.0) \;, \\
    p_3 &= p_1 - p_2 \;. \\
\end{split}
\end{equation*} 
The edge weights $\nu_e$ are set to $5/6$. We get the following values:
\begin{center}
\begin{tabular}{l c l l c l}
     $I_{\text{ref}}$ & $=$ & \num{9.76546e-5} & \\
         $I$ & $=$ & \num{9.76565(24)e-5} &$\Delta$ & $=$ & 
         $\num{0.918}\sigma, \; \num{0.002}\%$ \\
     $T_{\text{tr}}$ & $=$ & $\num{1.14} \unit{\us} $ & 
     $T_{\text{ltd}}$ & $=$ & $\num{745} \unit{\ns}$ \\ 
     $C^2$ & $=$ & $\num{0.604}$ &
\end{tabular}
\end{center}
The reference value $I_{\text{ref}}$ has been obtained analytically using OneLOop~\cite{vanHameren:2010cp}. 
Here, the time spent in sampling is longer than in evaluating the integrand. 
This is due to the simplicity of this particular topology.
We choose the value of $Q$ in~\eqref{eq:spherical_param} to be of the same order of magnitude as the external kinematics. It should be noted that further fine-tuning of the parameters $Q$ and $b$ could lead to faster convergence rates of the naive method. For this example, we set $Q=1.0$ and obtain following values: 
\begin{center}
\begin{tabular}{l c l l c l}
    $I_n$ & $=$ & $\num{9.7655(21)e-5}$ & $\Delta_n$ & $=$ & 
    $\num{0.038}\sigma_n, \; \num{0.001}\%$ \\
    $C_n^2$ & $=$ & $\num{46.2}$ &
\end{tabular}
\end{center}
We thus see that \texttt{momtrop} converges roughly 77 times faster than the naive implementation.

\subsection{Example: a 6-point 2-loop topology}
We increase the complexity of the problem by doubling both the number of externals and the loop count. The topology we integrate is given by: 
\begin{align} \begin{tikzpicture} \begin{feynman} \vertex(int7); \vertex[below=5cm of int7](int8); \vertex[below left=1.5cm and 2.04cm of int7](int1); \vertex[below left=3.5cm and 2.04cm of int7](int2); \vertex[below=1.5cm of int7](int3); \vertex[below=3.5cm of int7](int4); \vertex[below right=1.5cm and 2.04cm of int7](int5); \vertex[below right=3.5cm and 2.04cm of int7](int6); \vertex[left=of int1](ext1); \vertex[left=of int2](ext2); \vertex[right=of int3](ext3); \vertex[right=of int4](ext4); \vertex[right=of int5](ext5); \vertex[right=of int6](ext6); \diagram*[large]{ (int7) -- [--,line width=0.4mm, black] (int8), (int7) -- [--,line width=0.4mm, black, half left] (int8), (int7) -- [--,line width=0.4mm, black, half right] (int8), (ext1) -- [--,line width=0.4mm, black, momentum=\(p_1\)] (int1), (ext2) -- [--,line width=0.4mm, black, momentum=\(p_2\)] (int2), (ext3) -- [--,line width=0.4mm, black, momentum=\(p_4\)] (int3), (ext4) -- [--,line width=0.4mm, black, momentum=\(p_3\)] (int4), (ext5) -- [--,line width=0.4mm, black, momentum=\(p_5\)] (int5), (int6) -- [--,line width=0.4mm, black, momentum=\(p_6\)] (ext6), }; \end{feynman} \end{tikzpicture} \end{align}
The external momenta are set to:
\begin{equation*}
\begin{split}
        p_1 &= (0.2, 0.3, 0.5, 0.6), \\ 
        p_2 &= (-0.1, 0.7, 0.2, 0.1), \\ 
        p_3 &= (0.1, 0.5, -0.3, -0.4), \\ 
        p_4 &= (-0.3, 0.4, 0.5, 0.2), \\ 
        p_5 &= (-0.2, 0.3, 0.2, -0.5), \\
        p_6 &= p_1 + p_2 + p_3 + p_4 + p_5.
\end{split}
\end{equation*}
The edge weights $\nu_e$ are set to 5/9. We find the following results:
\begin{center}
\begin{tabular}{l c l l c l}
     $I_{\text{ref}}$ & $=$ &\num{1.1339(5)e-4}. & \\
         $I$ & $=$ & \num{1.13355(20)e-4} &$\Delta$ & $=$ & 
         $\num{1.798}\sigma, \; \num{0.031}\%$ \\
     $T_{\text{tr}}$ & $=$ & $\num{2.56} \unit{\us} $ & 
     $T_{\text{ltd}}$ & $=$ & $\num{2.85} \unit{\us}$ \\
     $C^2$ & $=$ & $\num{31.1}$ &
\end{tabular}
\end{center}
The reference value $I_{\text{ref}}$ has been obtained numerically using pySecDec~\cite{Borowka:2017idc}. Using the naive implementation with $Q=0.2$ in~\eqref{eq:spherical_param} yields the following results. %
\begin{center}
\begin{tabular}{l c l l c l}
    $I_n$ & $=$ & \num{1.127(20)e-4} &$\Delta_n$ & $=$ & 
    $\num{0.312}\sigma_n, \; \num{0.551}\%$ \\
    $C_n^2$ & $=$ & $\num{3.11e5}$ &
\end{tabular}
\end{center}
We thus see that the \texttt{momtrop} version requires roughly $10^4$ times fewer sample points than the naive implementation to achieve the same accuracy.

\subsection{Example: a 2-point 3-loop topology}
Increasing the number of loops from 2 to 3, we turn our attention to the 
``Mercedes'' topology:
\begin{align} \begin{tikzpicture} \begin{feynman} \vertex(int1); \vertex[above right = 2.0cm and 2.0cm of int1](int2); \vertex[below right = 2.0cm and 2.0cm of int1](int3); \vertex[below right = 2.0cm and 0.0cm of int2](int4); \vertex[below right = 0.0cm and 2.0cm of int4](int5); \vertex[left=of int1](ext1); \vertex[right=of int5](ext2); \diagram*[large]{ (int1) -- [--,line width=0.4mm, black, quarter left] (int2), (int1) -- [--,line width=0.4mm, black, quarter right] (int3), (ext1) -- [--,line width=0.4mm, black, momentum=\(p_1\)] (int1), (int5) -- [--,line width=0.4mm, momentum=\(p_2\), black] (ext2), (int2) -- [--,line width=0.4mm, black] (int4), (int3) -- [--,line width=0.4mm, black] (int4), (int4) -- [--,line width=0.4mm, black] (int5), (int2) -- [--,line width=0.4mm, black, quarter left] (int5), (int3) -- [--,line width=0.4mm, black, quarter right] (int5), }; \end{feynman} \end{tikzpicture} \end{align}
The external momenta are set to:
\begin{equation*}
\begin{split}
    p_1 &= (0,0,0,1),  \\
    p_2 &= p_1.
\end{split}
\end{equation*}
The edge weights $\nu_e$ are all set to $11/14$. We find the following values after integration:
\begin{center}
\begin{tabular}{l c l l c l}
     $I_{\text{ref}}$ & $=$ & \num{5.26647e-6} & \\
         $I$ & $=$ & \num{5.26644(12)e-6} &$\Delta$ & $=$ & 
         $\num{0.223}\sigma, \; \num{0.001}\%$ \\
     $T_{\text{tr}}$ & $=$ & $\num{2.48} \unit{\us} $ & 
     $T_{\text{ltd}}$ & $=$ & $\num{1.34} \unit{\us}$ \\
     $C^2$ & $=$ & $\num{0.519}$ &
\end{tabular}
\end{center}
The reference value $I_{\text{ref}}$ has been obtained from the analytical result using FORCER \cite{Ruijl:2017cxj}.
Running the naive implementation with $Q=1.0$ gives the following results:
\begin{center}
\begin{tabular}{l c l l c l}
    $I_n$ & $=$ & \num{5.42(11)e-6} &$\Delta_n$ & $=$ & 
    $\num{1.315}\sigma_n, \; \num{2.848}\%$ \\
    $C_n^2$ & $=$ & $\num{4.36e5}$ &
\end{tabular}
\end{center}
For this example, \texttt{momtrop} achieves again a drastic acceleration of the convergence speed. To achieve the same accuracy, the naive implementation needs $10^{6}$ times the number of sample points that \texttt{momtrop} needs. 

In order to compare our method to other numerical approaches for loop integration, we have performed the same integral using pySecDec. We compare the total execution time in CPU hours, measured on the same system equipped with an Intel Xeon W-2135. Our run with momtrop lasted a total of 1.22 CPU hours. From these 1.22 CPU hours about 3 seconds where spent in generating the CFF expression and the preprocessing stage described in \ref{sec:pre}. The rest of the computation time was spent in evaluating the $10^9$ samples.

The pySecDec run took up 0.43 CPU hours, and achieved a result with a slightly better error, $I=\num{5.266502(81)e-06}$. In contrast to our method, most of the runtime is taken up by the preprocessing stage. From these 0.43 CPU hours, about 44 seconds are spent evaluating the integrand. This fast convergence time after preprocessing can be most likely attributed to the fact that pySecDec can use Quasi-Monte Carlo methods, which have errors proportional to $\frac{1}{N}$.

This comparison illustrates both the strengths and weaknesses of the tropical approach: the tropical sampling method requires minimal preprocessing, enabling quick acquisition of low-accuracy results. This makes it particularly well-suited for scenarios where many evaluation points are needed but only moderate precision is sufficient. Conversely, when high-accuracy results are required, other methods tend to outperform the tropical approach, unless the tropical sampling implementation is enhanced, for example, by incorporating quasi-Monte Carlo techniques on top of tropical sampling to improve convergence rates.

\subsection{Example: a 2-point 4-loop topology}
The following test is a 2-point, 4-loop non-planar 
topology: 
\begin{align} \begin{tikzpicture} \begin{feynman} \vertex(int1); \vertex[above right=2cm and 2cm of int1](int2); \vertex[right=4cm of int1](int3); \vertex[below=4cm of int2](int4); \vertex[left=1cm of int3](int5); \vertex[below=1cm of int2](int6); \vertex[left=1.9cm of int3](int7); \vertex[right=1.9cm of int1](int8); \vertex[right=of int3](ext2); \vertex[below right=1.1cm and 0.35cm of int1](int9); \vertex[left=of int9](ext1); \diagram*[large]{ (int1) -- [--,line width=0.4mm, black, quarter left] (int2), (int2) -- [--, black, line width=0.4mm, quarter left] (int3), (int3) -- [--, black, line width=0.4mm, quarter left] (int4), (int4) -- [--, black, line width=0.4mm, quarter left] (int1), (int2) -- [--, line width=0.4mm, black] (int4), (int3) -- [--, line width=0.4mm, black] (int5), (int5) -- [--, line width=0.4mm, black] (int6), (int5) -- [--, line width=0.4mm, black] (int7), (int8) -- [--, line width=0.4mm, black] (int1), (int3) -- [--, line width=0.4mm, black, momentum=\(p_2\)] (ext2), (ext1) -- [--,line width=0.4mm, black, momentum=\(p_1\)] (int9), }; \end{feynman} \end{tikzpicture} \end{align}
The external momenta are set to 
\begin{equation*}
\begin{split}
    p_1 &= (0,0,0,1),  \\
    p_2 &= p_1.
\end{split}
\end{equation*}
The edge weights $\nu_e$ are set to $7/9$. We find the following results:
\begin{center}
\begin{tabular}{l c l l c l}
     $I_{\text{ref}}$ & $=$ & \num{8.36515e-8} & \\
         $I$ & $=$ & \num{8.36522(24)e-8} &$\Delta$ & $=$ & 
         $\num{0.275}\sigma, \; \num{0.001}\%$ \\
     $T_{\text{tr}}$ & $=$ & $\num{2.54} \unit{\us} $ & 
     $T_{\text{ltd}}$ & $=$ & $\num{1.50} \unit{\us}$ \\ 
     $C^2$ & $=$ & $\num{0.823}$ 
\end{tabular}
\end{center}
The reference value $I_{\text{ref}}$ has been obtained from the analytical result using FORCER \cite{Ruijl:2017cxj}.
Setting $Q=1.0$, the naive implementation yields the following results:
\begin{center}
\begin{tabular}{l c l l c l}
    $I_n$ & $=$ & \num{8.73(46)e-8} &$\Delta_n$ & $=$ & 
    $\num{0.788} \sigma_n, \; \num{4.358}\%$ \\
    $C_n^2$ & $=$ & $\num{3.02e6}$ &
\end{tabular}
\end{center}
This is the most drastic improvement in convergence speed we have observed so far. With tropical sampling, \num{3.7e6} times fewer sample points are required to obtain the same level of accuracy as the naive method. 

\subsection{Example: a 4-point 4-loop topology}
Pushing the complexity further, we integrate the following $2 \cross 2$ fishnet topology:
\begin{align} \begin{tikzpicture} \begin{feynman} \vertex(int1); \vertex[right=2cm of int1](int2); \vertex[right=2cm of int2](int3); \vertex[below=2cm of int1](int4); \vertex[below=2cm of int2](int5); \vertex[below=2cm of int3](int6); \vertex[below=2cm of int4](int7); \vertex[below=2cm of int5](int8); \vertex[below=2cm of int6](int9); \vertex[left=of int1](ext1); \vertex[right=of int3](ext2); \vertex[left=of int7](ext3); \vertex[right=of int9](ext4); \diagram*[large]{ (int1) -- [--,line width=0.4mm, black] (int2), (int2) -- [--,line width=0.4mm, black] (int3), (int4) -- [--,line width=0.4mm, black] (int5), (int5) -- [--,line width=0.4mm, black] (int6), (int1) -- [--,line width=0.4mm, black] (int4), (int2) -- [--,line width=0.4mm, black] (int5), (int3) -- [--,line width=0.4mm, black] (int6), (int7) -- [--,line width=0.4mm, black] (int8), (int9) -- [--,line width=0.4mm, black] (int8), (int4) -- [--,line width=0.4mm, black] (int7), (int5) -- [--,line width=0.4mm, black] (int8), (int6) -- [--,line width=0.4mm, black] (int9), (ext1) -- [--,line width=0.4mm, black, momentum=\(p_1\)] (int1), (ext3) -- [--,line width=0.4mm, black, momentum=\(p_2\)] (int7), (int3) -- [--,line width=0.4mm, black, momentum=\(p_3\)] (ext2), (int9) -- [--,line width=0.4mm, black, momentum=\(p_4\)] (ext4), }; \end{feynman} \end{tikzpicture} \end{align}
The external momenta are set to 
\begin{equation*}
\begin{split}
    p_1 &= (2, -5.2, 2.1, 0.0), \\
    p_2 &= (1.2, 2.2, 1, 0.4), \\ 
    p_3 &= (1.6, -0.1, 12.5, -2.4), \\ 
    p_4 &= p_1 + p_2 - p_3.
\end{split}
\end{equation*}
The edge weights $\nu_e$ are set to $3/4$. We find the following results:

\begin{center}
\begin{tabular}{l c l l c l}
     $I_{\text{ref}}$ & $=$ &\num{2.6919e-14} & \\
     $I$ & $=$ & \num{2.6875(23)e-14} &$\Delta$ & $=$ & 
     $\num{1.915}\sigma, \; \num{0.161}\%$ \\
     $T_{\text{tr}}$ & $=$ & $\num{2.92} \unit{\us} $ & 
     $T_{\text{ltd}}$ & $=$ & $\num{16.27} \unit{\us}$ \\ 
     $C^2$ & $=$ & $\num{730}$
\end{tabular}
\end{center}
The reference value of this integral is taken 
from the analytical result \cite{Basso:2017jwq}. The naive implementation with $Q=3.0$ gives the following results:
\begin{center}
\begin{tabular}{l c l l c l}
    $I_n$ & $=$ & \num{3.32(59)e-14} &$\Delta_n$ & $=$ & 
    $\num{1.049}\sigma_n, \; \num{23.2}\%$ \\
    $C_n^2$ & $=$ & $\num{4.80e7}$ &
\end{tabular}
\end{center}
In this case, \texttt{momtrop} gives yet another big improvement in the convergence speed, requiring around \num{6.6e4} times fewer sample points than the naive method. 

\subsection{Example: a 4-point 6-loop topology}
Pushing LTD to its limits, we integrate the 
$2 \cross 3$ fishnet topology:
\begin{align} \begin{tikzpicture} \begin{feynman} \vertex(int1); \vertex[right=2cm of int1](int2); \vertex[right=2cm of int2](int3); \vertex[right=2cm of int3](int4); \vertex[below=2cm of int1](int5); \vertex[below=2cm of int2](int6); \vertex[below=2cm of int3](int7); \vertex[below=2cm of int4](int8); \vertex[below=2cm of int5](int9); \vertex[below=2cm of int6](int10); \vertex[below=2cm of int7](int11); \vertex[below=2cm of int8](int12); \vertex[left=of int1](ext1); \vertex[left=of int9](ext2); \vertex[right=of int4](ext3); \vertex[right=of int12](ext4); \diagram*[large]{ (int1) -- [--,line width=0.4mm, black] (int2), (int2) -- [--,line width=0.4mm, black] (int3), (int3) -- [--,line width=0.4mm, black] (int4), (int1) -- [--,line width=0.4mm, black] (int5), (int2) -- [--,line width=0.4mm, black] (int6), (int3) -- [--,line width=0.4mm, black] (int7), (int4) -- [--,line width=0.4mm, black] (int8), (int5) -- [--,line width=0.4mm, black] (int6), (int6) -- [--,line width=0.4mm, black] (int7), (int7) -- [--,line width=0.4mm, black] (int8), (int5) -- [--,line width=0.4mm, black] (int9), (int6) -- [--,line width=0.4mm, black] (int10), (int7) -- [--,line width=0.4mm, black] (int11), (int8) -- [--,line width=0.4mm, black] (int12), (int9) -- [--,line width=0.4mm, black] (int10), (int10) -- [--,line width=0.4mm, black] (int11), (int11) -- [--,line width=0.4mm, black] (int12), (ext1) -- [--,line width=0.4mm, black, momentum=\(p_1\)] (int1), (ext2) -- [--,line width=0.4mm, black, momentum=\(p_2\)] (int9), (int4) -- [--,line width=0.4mm, black, momentum=\(p_3\)] (ext3), (int12) -- [--,line width=0.4mm, black, momentum=\(p_4\)] (ext4), }; \end{feynman} \end{tikzpicture} \end{align}
The external momenta are set to 
\begin{equation*}
\begin{split}
    p_1 &= (2, -5.2, 2.1, 0.0), \\
    p_2 &= (1.2, 2.2, 1, 0.4), \\ 
    p_3 &= (1.6, -0.1, 12.5, -2.4), \\ 
    p_4 &= p_1 + p_2 - p_3.
\end{split}
\end{equation*}
The edge weights $\nu_e$ are set to 12/17. The following results are 
obtained with $N=\num{2.136e9}$ sample points: 
\begin{center}
\begin{tabular}{l c l l c l}
     $I_{\text{ref}}$ & $=$ &\num{8.4045e-19} & \\
         $I$ & $=$ & \num{8.4089(98)e-19} &$\Delta$ & $=$ & 
         $\num{0.453}\sigma, \; \num{0.053}\%$ \\
     $T_{\text{tr}}$ & $=$ & $\num{15.39} \unit{\us} $ & 
     $T_{\text{ltd}}$ & $=$ & $\num{1.94} \unit{\ms}$ \\ 
     $C^2$ & $=$ & $\num{2.90e3}$ 
\end{tabular}
\end{center}
The reference value $I_{\text{ref}}$ is taken from the analytical result \cite{Basso:2017jwq}.
For this example, we run the naive implementation with $Q=3.0$. However, this run fails to converge to the correct 
value:
\begin{center}
\begin{tabular}{l c l l c l}
    $I_n$ & $=$ & \num{2.67(78)e-19 } &$\Delta_n$ & $=$ & 
    $\num{7.342}\sigma_n, \; \num{68.2}\%$ \\
\end{tabular}
\end{center}
While the naive implementation fails to give a reliable result for this highly complex example, \texttt{momtrop} achieves a very accurate result given the same number of samples.

\subsection{Example: a 6-photon 1-loop topology}
To illustrate that this method still works with massive propagators and physical numerators, we integrate the following 6-photon diagram with a top quark in the loop.
\begin{align} \begin{tikzpicture} \begin{feynman} \vertex(int1); \vertex[above=4cm of int1](int2); \vertex[right=4cm of int2](int3); \vertex[below=1.33cm of int3](int4); \vertex[below=1.33cm of int4](int5); \vertex[below=1.33cm of int5](int6); \vertex[left=of int1](ext1); \vertex[left=of int2](ext2); \vertex[right=of int3](ext3); \vertex[right=of int4](ext4); \vertex[right=of int5](ext5); \vertex[right=of int6](ext6); \diagram*[large]{ (int1) -- [fermion,line width=0.4mm, black] (int2), (int2) -- [fermion,line width=0.4mm, black] (int3), (int3) -- [fermion,line width=0.4mm, black] (int4), (int4) -- [fermion,line width=0.4mm, black] (int5), (int5) -- [fermion,line width=0.4mm, black] (int6), (int6) -- [fermion,line width=0.4mm, black] (int1), (ext1) -- [photon,line width=0.4mm, black, momentum=\(p_1\)](int1), (ext2) -- [photon,line width=0.4mm, black, momentum=\(p_2\)](int2), (int3) -- [photon,line width=0.4mm, black, momentum=\(p_3\)](ext3), (int4) -- [photon,line width=0.4mm, black, momentum=\(p_4\)](ext4), (int5) -- [photon,line width=0.4mm, black, momentum=\(p_5\)](ext5), (int6) -- [photon,line width=0.4mm, black, momentum=\(p_6\)](ext6), }; \end{feynman} \end{tikzpicture} \end{align}
The kinematics are set to 
\begin{equation*}
\begin{split}
    p_1 &= (500, 0, -300, 400), \\
    p_2 &= (500, 0, 300, -400), \\ 
    p_3 &= (88.551333054502976,-22.100690287689979, \\ 
    &40.080353191685333,-75.805430956936632). \\ 
    p_4 &= (328.32941922709853,-103.84961188345630, \\
    &-301.93375538954012,76.494921387165888) \\ 
    p_5 &= (152.35810946743061,-105.88095966659220, \\ 
    &-97.709638326975707,49.548385226792817
\end{split}
\end{equation*}
The edge weights are set to $\nu_e = 5/12$. The top mass is set to $m_t = 1500 \; \unit{\GeV}$ in order to remain below threshold. The electroweak 
coupling is set to $\alpha_{EW} = 1/128.93$. The helicities of 
the external photons are set to $(-,-,-,-,-,-)$. We find the following results
\begin{center}
\begin{tabular}{l c l l c l}
     $I_{\text{ref}}$ & $=$ & $\num{1.22898e-13} + \num{3.94362e-13}i$& \\
         $I$ & $=$ & $\num{1.22894(12)e-13} + \num{3.94364(13)e-13} i $ \\
         $\Delta_{\text{real}}$ & $=$ & $\num{0.374}\sigma, \; \num{0.004}\%$ \\
         $\Delta_{\text{imag}}$ & $=$ & $\num{0.118}\sigma, \; \num{0.000}\%$ \\
     $T_{\text{tr}}$ & $=$ & $\num{1.89} \unit{\us} $ \\ 
     $T_{\text{ltd}}$ & $=$ & $\num{17.65} \unit{\us}$ \\
     $C_{\text{real}}^2$ & $=$ & $\num{9.53}$ \\
     $C_{\text{imag}}^2$ & $=$ & $\num{1.09}$ 
\end{tabular}
\end{center}
The reference value $I_{\text{ref}}$ has been obtained from the analytical result with MadLoop~\cite{Hirschi:2011pa}.
We also perform the integral with the naive implementation. Setting $Q = 1000.0$ yields the following results:
\begin{center}
\begin{tabular}{l c l l c l}
         $I_n$ & $=$ & $\num{1.22923(19)e-13} + \num{3.94390(24)e-13} i $ \\
         $\Delta_{\text{real},n}$ & $=$ & $\num{1.270}\sigma_n, \; \num{0.020}\%$ \\
         $\Delta_{\text{imag},n}$ & $=$ & $\num{1.163}\sigma_n, \; \num{0.007}\%$ \\
     $C_{\text{real},n}^2$ & $=$ & $\num{23.9}$ \\
     $C_{\text{imag},n}^2$ & $=$ & $\num{3.70}$ 
\end{tabular}
\end{center}
For this example \texttt{momtrop} provides a relatively minor improvement over the naive implementation. This can be explained by the fact that the naive implementation is free of integrable singularities due to the top mass.

\section{Performance comparison to multi-channelling methods}
\label{sec:multi}
In this section, we briefly compare our method to the multi-channeling approach, which provides an alternative method to deal with integrable singularities \cite{Becker:2012aqa, Capatti:2019edf}. The basic idea of this method is to split the integrand into different \emph{channels}, such that each channel contains a single singularity at the origin, which a spherical-like coordinate transformation can remove. 

As long as the number of edges remains moderate (i.e., fewer than approximately 20), the preprocessing step described in Section~\ref{sec:pre} incurs negligible computational cost. Only at very high loop orders do these costs become significant and eventually dominant~\cite{Balduf:2023ilc, Borinsky:2025ywo}. Importantly, the preprocessing step does not depend on the exact numerical values of the masses and external momenta, as long as the qualitative structure of the kinematic configuration remains unchanged, meaning masses or momenta being zero or nonzero. As a result, a single preprocessing step suffices to handle broad ranges of mass and momentum values without the need for recomputation.

In the fully massless case of an $L$ loop graph, at most $L$ propagators can vanish simultaneously. A set of $L$ propagators may vanish simultaneously if the complement of their associated edges forms a spanning tree $T$ of the underlying graph $G$. We divide the integrand into 
[number of spanning trees] many
channels by inserting the identity
\begin{equation}
    1 = \frac{1}{\sum\limits_{T \subset G} \prod\limits_{e \notin T} 
    D_e^{-\alpha/2}} 
    \sum\limits_{T \subset G} \prod\limits_{e \notin T} 
    D_e^{-\alpha/2}\;,
\label{eq:multi-channeling}
\end{equation}
where $\alpha$ is a tunable parameter. Each term in the right sum in Eq.~\eqref{eq:multi-channeling} corresponds to a channel which we parameterize separately. The sum in the denominator appears in each channel and ensures that only the propagators in the numerator constitute an integrable singularity. For each channel, we change the momentum routing such that $\prod_{e \notin T} D_e = \prod_{\ell} (k_\ell^2) \; .$ Parameterizing each loop momenta in a spherical coordinate system centred at the origin, we obtain a factor $\prod_{\ell} (k_\ell^2)^{(D-1)/2}$, which may cancel integrable singularities. In our implementation, we have opted to implement the sum over channels in a Monte Carlo fashion, by uniformly sampling a channel for each set of sampled loop momenta.

We compare three integration methods using $N=10^9$ sample points for the Mercedes topology. The first integration method uses parameterization~\eqref{eq:spherical_param} without multi-channelling or tropical sampling. The parameter $b$ is set to $b=10$. The second integration method uses the same parameterization and multi-channelling with $\alpha=2$. The final integration method uses tropical sampling with edge weights $\nu_e = 11/14$.
\begin{center}
\begin{tabular}{l|l|c}
     & $I$ & $C^2$ \\
     \hline
     Spherical Parameterization & $\num{5.42(11)e-6}$ & $\num{4.36e5}$ \\
     Multi-channeling & $\num{5.2656(69)e-6 }$ & $\num{1.72e3}$ \\
     Tropical sampling & $\num{5.26644(12)e-6}$ & $\num{0.519}$
\end{tabular}
\end{center}

We see that the multi-channelling procedure gives a factor of 250 improvement over the naive method. The tropical sampling method, however, converges roughly $10^6$ times faster than the naive approach. While the multi-channeling approach successfully removes integrable singularities, the complicated prefactor in Eq.~\eqref{eq:multi-channeling} introduces new structures that are not absorbed
into the sampling measure. Since tropical sampling does not need such prefactors, it can achieve far better convergence rates.

The CFF expressions used in the benchmarks are generated by $\gamma$Loop and converted to \verb|C++| and inline assembly using \verb|Symbolica| (\url{https://symbolica.io/}). We have included all factors of $i$, such that scalar integrals evaluate to positive real numbers for Euclidean kinematics.

\section{Conclusion \& Outlook}
\label{sec:con}

In this paper, we introduced a new algorithm to sample points distributed as the integrand of a sufficiently well-behaved Feynman integral. We call the associated probability measure the \emph{Feynman measure} and our algorithm constitutes the first systematic method to sample from this measure. In combination with a Monte Carlo pipeline, our algorithm can be used to evaluate Feynman and phase-space integrals. We implemented this algorithm as the \texttt{Rust} software package \texttt{momtrop}. This package can be used as a standalone component to produce samples from the Feynman measure. 

We illustrated the capabilities of both the algorithm and the software package via visualizations and by running benchmarks against naive evaluation techniques for phase-space type integrals. The achieved speedup factors of $10^{6}$ and more for nontrivial topologies show that our approach puts regimes within reach which were inaccessible using previous methods.

Older evaluation methods for phase-space-type integrals typically use naive evaluation techniques combined with black-box sampling adaptation methods, such as \texttt{VEGAS} \cite{Lepage:1977sw}, to reduce the sampling variance. Our benchmark comparisons are, hence, not entirely fair, as we do not combine the naive method with any adaptation method. On the other hand, we could still easily put adaptive sampling measures on top of our tropical sampling algorithm to further reduce the variance. For instance, this would be relevant for loop-tree-duality integrals in gauge theories, where the integration kernel $f(\bb k)$ is not necessarily positive. Systematically optimizing the spitting explained at the beginning of Section~\ref{sec:ltd} constitutes another potential source of variance reduction for our approach.

Another possible extension of our approach is to add further aspects of the integration kernel $f(\bb k)$ to the sampling process. This is particularly promising in situations where $f(\bb k)$ exhibits (integrable) singularities. Our algorithm in Section~\ref{sec:algo} is the first of its kind as a dedicated method to sample from Feynman-type measures. We expect further tweaks of the method to lead to more refined and optimized sampling procedures.

We performed these benchmarks and verified the validity of our algorithm on integrals that come from the loop-tree-duality framework. The \texttt{momtrop} package is also available as a subcomponent of the $\gamma$Loop software package, which evaluates loop-tree-duality integrals.

We anticipate further applications of our algorithm to phase-space integrals relevant for collider phenomenology. Evaluating these integrals constitutes a severe bottleneck within collider physics phenomenology workflows. Further, a key bottleneck within the promising local unitarity approach towards cross-section computations is the numerical stability of the integration algorithms. We expect our approach, which we have already tested within the loop-tree-duality framework, to also be highly beneficial in this domain.

\section*{Acknowledgments}
We thank Valentin Hirschi for valuable comments on an early version of the manuscript. We are grateful to Babis Anastasiou, Zeno Capatti, Thomas Gehrmann, Valentin Hirschi, Lucien Huber,  William J. Torres Bobadilla, and Paolo Torrielli for inspirational discussions during various stages of this project. MF also thanks Valentin Hirschi for general support.
Research at Perimeter Institute is supported in part by the Government of Canada through the Department of Innovation, Science and Economic Development and by the Province of Ontario through the Ministry of Colleges and Universities. MB was partially supported by Dr.\ Max Rössler, the Walter Haefner Foundation and the ETH Zürich Foundation. MF was supported by the SNSF grant PCEFP2 203335.

\appendix

\section{Worked out example of the tropical sampling algorithm}
\label{ap:example}

To provide some intuition on the algorithm's inner workings described in Section~\ref{sec:samp}, we take a closer look at some of the key steps of this sampling strategy. We will do so by studying the double triangle topology. We will also give explicit expressions for some defined quantities for increased concreteness. 

Let $G$ be the graph for the double triangle diagram, which we will consider as a set of edges
\begin{align*} G &= \resizebox{5cm}{!}{\raisebox{-0.46cm}{ \begin{tikzpicture} \begin{feynman} \vertex(int1); \vertex[above right = 0.4cm and 0.6cm of int1](int2); \vertex[below right = 0.4cm and 0.6cm of int1](int3); \vertex[below right = 0.4cm and 0.6cm of int2](int4); \vertex[left = 0.2cm of int1](ext1); \vertex[right = 0.2cm of int4](ext4); \vertex[above right = 0.15cm and 0.03cm of int1](L1){\resizebox{0.15cm}{!}{$e_1$}}; \vertex[below right = 0.15cm and 0.03cm of int1](L2){\resizebox{0.15cm}{!}{$e_2$}}; \vertex[above right = 0.23cm and -0.07cm of int3](L3){\resizebox{0.15cm}{!}{$e_3$}}; \vertex[above left = 0.15cm and 0.03cm of int4](L4){\resizebox{0.15cm}{!}{$e_4$}}; \vertex[below left = 0.15cm and 0.03cm of int4](L5){\resizebox{0.15cm}{!}{$e_5$}}; \diagram*[large]{ (int1) -- [--,line width=0.2mm, black] (int2), (int1) -- [--,line width=0.2mm, black] (int3), (int2) -- [--,line width=0.2mm, black] (int3), (int2) -- [--,line width=0.2mm, black] (int4), (int3) -- [--,line width=0.2mm, black] (int4), (ext1) -- [--, line width=0.2mm, black](int1), (int4) -- [--, line width=0.2mm, black](ext4), }; \end{feynman} \end{tikzpicture} }} = \{e_1, e_2, e_3, e_4, e_5 \} \;. \end{align*}
Each edge $e$ of the graph is associated with a Feynman parameter $x_e$ and an edge weight $\nu_e$. We choose edges 2 and 3 to have a mass $m$, and we assume an external momentum $p$ flows in on the left side of the graph and out on the right side. We can use the definitions for $\mathcal{U}$ and $\mathcal{F}$ given in Section~\ref{sec:bounds} to determine the first and second Symanzik polynomials for the graph $G$:
\begin{equation*}
\begin{split}
    \mathcal{U}(x) &= x_1 x_3 + x_1 x_4 + x_1 x_5 + x_2 x_3 + x_2 x_4 + x_2 x_5 
    + x_3 x_4 + x_3 x_5 \\ 
    \mathcal{F}(x) &= p^2 x_2 x_3 x_4 + p^2 x_1 x_3 x_5 + p^2 x_1 x_2 x_3 + p^2 x_3 x_4 x_5 \\
    & \qquad \qquad +  p^2 x_2 x_4 x_5 + p^2 x_1 x_4 x_5 + p^2 x_1 x_2 x_5 + p^2 x_1 x_2 x_4  + m^2 \; \mathcal{U}(x) 
    (x_2 + x_3) \;.
\end{split}
\end{equation*}
Let us turn our attention to the integral over the tropical measure, given by
\begin{equation}
\label{eq:mu_tr_dt}
        \int \mu^{\tr} = \frac{1}{I^{{\tr}}} \int_{x_e \geq 0} \left(\prod_e  \frac{\dd{x_e}}{x_e} \right ) \delta\left(1-\sum_e \rho_e x_e\right)
        \frac{x_1^{\nu_1} x_2^{\nu_2} x_3^{\nu_3} 
    x_4^{\nu_4} x_5^{\nu_5} }
    { \mathcal{U}^{\tr}(x)^{D/2} \mathcal{V}^{\tr}(x)^{\omega}},
\end{equation}
where, by the Cheng--Wu theorem, the $\rho_e$ can be any set of non-negative numbers, not all $0$.
We divide the domain of integration into $5! = 120$ sectors, each defined by an ordering of Feynman parameters. Let us focus on the sector defined by the ordering $x_1 > x_2 > x_3 > x_4 > x_5$. This sector will be denoted by $\sigma$. An ordering of Feynman parameters induces a partial order on the sets of monomials of $\mathcal{U}$ and $\mathcal{F}$. If there is a well-defined maximum under this partial order, we can use this to determine an explicit monomial that corresponds to  $\mathcal{U}^{\tr}$ and $\mathcal{F}^{\tr}$ in that particular sector. For the sector $\sigma$, we find
\begin{equation*}
\begin{split}
    \mathcal{U}^{\tr}(x) &= x_1 x_3 \;, \\
    \mathcal{F}^{\tr}(x) &= x_1 x_2 x_3 \;, \\
    \mathcal{V}^{\tr}(x) &= \frac{\mathcal{F}^{\tr}(x)}{\mathcal{U}^{\tr}(x)} = x_2 \;.
\end{split}
\end{equation*}

Since $\mathcal{U}^{\tr}$ and $\mathcal{V}^{\tr}$ take simple forms when restricting to the domain $\sigma$, we can easily compute the contribution of $\int \mu_{\tr}$ coming from $\sigma$ by integrating over the Feynman parameters.
We first use our freedom to freely choose the $\rho$ parameters to set
$\rho_1 =1$ and $\rho_2=\rho_3=\rho_4=\rho_5=0$ and therefore $x_1=1$ resolving the $\delta$ function. So,
\begin{equation}
\label{eq:integrating}
\begin{split}
    \int_{\sigma} \mu^{\tr} &= \frac{1}{I^{\tr}} \int_0^{1} \dd{x_2} 
    \int_0^{x_2} \dd{x_3} 
    \int_{0}^{x_3} \dd{x_4} \int_0^{x_4} \dd{x_5} \frac{x_2^{\nu_2 
    - 1} x_3^{\nu_3 - 1} 
    x_4^{\nu_4 - 1} x_5^{\nu_5 - 1} }
    { x_3^{D/2} x_2^{\omega}} \\
    &= \frac{1}{I^{\tr}}  \int_0^{1} \dd{x_2} \int_0^{x_2} 
    \dd{x_3} 
    \int_{0}^{x_3} \dd{x_4} \int_0^{x_4} \dd{x_5}\; x_2^{\nu_2 
    - \omega - 1} x_3^{\nu_3 - D/2 - 1} 
    x_4^{\nu_4 - 1} x_5^{\nu_5 - 1} \\
    &= \frac{1}{I^{\tr}} \frac{1}{\nu_5} \int_0^{1} \dd{x_2} 
    \int_0^{x_2} \dd{x_3} 
    \int_{0}^{x_3} \dd{x_4} \; x_2^{\nu_2 
    - \omega - 1} x_3^{\nu_3 - D/2 - 1} 
    x_4^{\nu_4 + \nu_5 - 1} \\ 
    &= \frac{1}{I^{\tr}} \frac{1}{\nu_5} \frac{1}{\nu_4 + \nu_5}  
    \int_0^{1} \dd{x_2} \int_0^{x_2} \dd{x_3} \;
    x_2^{\nu_2 - \omega - 1} x_3^{\nu_3 + \nu_4 + \nu_5 - D/2 - 1}  \\ 
    &= \frac{1}{I^{\tr}} \frac{1}{\nu_5} \frac{1}{\nu_4 + \nu_5} 
    \frac{1}{\nu_3 + \nu_4 + \nu_5 - D/2 } 
    \frac{1}{\nu_2 + \nu_3 + \nu_4 + \nu_5 - D/2} \; .
\end{split}
\end{equation}
The denominators above can be reexpressed using the generalized degree of divergence (Eq.~\eqref{eq:gen_dod}):
\begin{equation*}
    \int \mu^{\tr} = \frac{1}{I^{\tr}} \frac{1}{\omega(\{e_5\})} 
    \frac{1}{\omega(\{e_5, e_4\})} \frac{1}{\omega(\{e_5, e_4, e_3\})} 
    \frac{1}{\omega(\{e_5, e_4, e_3, e_2\})} \;.
\end{equation*}
Note that if a proper subgraph $\gamma$ satisfies $\omega(\gamma) \leq 0$, then the steps performed in Eq.~\eqref{eq:integrating} are no longer valid. In such cases, the integral has a non-integrable singularity at the integration boundary. This is expected, since $\omega(\gamma) \leq 0$ identifies UV or soft IR sub-divergences. 

The structure suggested by Eq.~\eqref{eq:integrating} allows importance sampling over sectors. Recall that in Section~\ref{sec:samp}, we iteratively remove edges from the graph $G$ with the probability of removing a specific edge from the subgraph $\gamma$ given by:
\begin{equation*}
    p_{\gamma}(e) = \frac{1}{J(\gamma)} \frac{J(\gamma \backslash e)}{\omega(\gamma \backslash e)} \;.
\end{equation*}
The order in which the edges get removed corresponds exactly to a sector. This can be seen by computing the probability of removing the edges of $G$ in the order $e_1, e_2, e_3, e_4, e_5$. This probability is given by the following product:
\begin{equation*}
\begin{split}
     p_G(e_1) &\cdot p_{G \backslash e_1} (e_2) \cdot p_{G \backslash \{e_1, e_2\}}(e_3)
    \cdot p_{G \backslash \{e_1, e_2, e_3\}}(e_4) \cdot p_{G \backslash \{e_1, e_2, e_3, 
    e_4\}}(e_5) \\
   &= \frac{1}{J(G)} \frac{J(G \backslash e_1)}{\omega(G \backslash e_1)} 
   \frac{1}{J(G \backslash e_1 )} \frac{J(G \backslash \{e_1, e_2\} )}{\omega(G 
   \backslash \{e_1, e_2\})} 
   \frac{1}{J(G \backslash \{e_1, e_2\} )} \frac{J(G \backslash \{e_1, e_2, e_3\} 
   )}{\omega(G \backslash \{e_1, e_2, e_3\})} \\ &\frac{1}{J(G \backslash \{e_1, 
   e_2, e_3 \} )} \frac{J(G \backslash \{e_1, e_2, e_3, e_4\} )}{\omega(G 
   \backslash \{e_1, e_2, e_3, e_4\})} 
   \frac{1}{J(G \backslash \{e_1, e_2, e_3, e_4 \} )} \frac{J(\emptyset)}
   {\omega(\emptyset)} \\ 
   &= \frac{1}{J(G)} \frac{1}{\omega(G \backslash e_1)} \frac{1}{\omega(G 
   \backslash \{e_1, e_2\})}
    \frac{1}{\omega(G \backslash \{e_1, e_2, e_3\})} \frac{1}{\omega(G \backslash 
    \{e_1, e_2, e_3, 
    e_4\})} \\ &= 
    \frac{1}{I^{\tr}} \frac{1}{\omega(\{e_5, e_4, e_3, e_2\})} 
    \frac{1}{\omega(\{e_5, e_4, e_3\})} 
    \frac{1}{\omega(\{e_5, e_4\})}  \frac{1}{\omega(\{e_5\})} \;.
\end{split}
\end{equation*}
If we identify $J(G) = I^{\tr}$, we see that this corresponds exactly to the quantity $\int_\sigma \mu^{\tr}$ that we computed earlier.

It remains to sample a set of Feynman parameters in a particular sector. Let us consider the quantity $\int_\sigma \mu^{\tr} f(x)$, where $f(x)$ is an arbitrary function that depends on all Feynman parameters. In our example, we have:
\begin{equation*}
  \int_\sigma \mu^{\tr} f(x) = \frac{1}{I^{\tr}} \int_0^{1} \dd{x_2} 
    \int_0^{x_2} \dd{x_3} 
    \int_{0}^{x_3} \dd{x_4} \int_0^{x_4} \dd{x_5} \; x_2^{\nu_2 
    - \omega - 1} x_3^{\nu_3 - D/2 - 1} 
    x_4^{\nu_4 - 1} x_5^{\nu_5 - 1} f(x) \;.
\end{equation*}
The powers of the Feynman parameters may be re-expressed in terms of the $\omega$'s:
\begin{equation*}
\begin{split}
  \int \mu^{\tr} &= \frac{1}{I^{\tr}} \int_0^{1} \dd{x_2} 
    \int_0^{x_2} \dd{x_3} 
    \int_{0}^{x_3} \dd{x_4} \int_0^{x_4} \dd{x_5} \\ 
    &\cross x_2^{ \omega(G \backslash 
    \{e_1\}) - \omega(G \backslash \{e_1, e_2\}) - \omega(G \backslash \{e_1, 
    e_2, e_3\}) 
    - \omega(G \backslash \{e_1, e_2, e_3, 
    e_4 \}) - 1} \\ 
    &\cross x_3^{\omega(G \backslash \{e_1, e_2\}) - \omega(G 
    \backslash 
    \{e_1, e_2, e_3\}) - \omega(G \backslash \{e_1, e_2, e_3, 
    e_4 \}) - 1} \\
    &\cross x_4^{\omega(G \backslash \{e_1, e_2, e_3\}) - \omega(G \backslash \{e_1, 
    e_2, e_3, 
    e_4 \}) - 1} \\
    &\cross x_5^{\omega(G \backslash \{e_1, e_2, e_3, e_4 \}) - 1} f(x) \;.
\end{split}
\end{equation*}
Now, we can perform the following coordinate transformation to `flatten' the powers of $x_e$ in front of $f(x)$: 
\begin{equation}
\begin{split}
\label{eq:transform}
    x_1 &= 1 \;, \\
    x_2 &= \xi_2^{ 1 / \omega(G \backslash \{e1\})} \;, \\
    x_3 &= x_2  \xi_3^{1 / \omega(G \backslash \{e1, e2\})} \;,  \\
    x_4 &= x_3 \xi_4^{1 / \omega(G \backslash \{e1, e2, e3\})} \;, \\
    x_5 &= x_4 \xi_5^{1 / \omega(G \backslash \{e1, e2, e3, e4\})} \;.
\end{split}
\end{equation}
Plugging this in, starting from $x_5$: 
\begin{equation*}
\begin{split}
  \int \mu^{\tr} &= \frac{1}{I^{\tr}} \frac{1}{\omega(G \backslash \{e_1, e_2, e_3,  e_4 \}} \int_0^{1} \dd{x_2} 
    \int_0^{x_2} \dd{x_3} 
    \int_{0}^{x_3} \dd{x_4} \int_0^{1} \dd{\xi_5} \\ & \qquad \cross x_2^{ \omega(G \backslash 
    \{e_1\}) - \omega(G \backslash \{e_1, e_2\}) - \omega(G \backslash \{e_1, 
    e_2, e_3\}) 
    - \omega(G \backslash \{e_1, e_2, e_3, 
    e_4 \}) - 1} \\ & \qquad \cross  x_3^{\omega(G \backslash \{e_1, e_2\}) - \omega(G 
    \backslash 
    \{e_1, e_2, e_3\}) - \omega(G \backslash \{e_1, e_2, e_3, 
    e_4 \}) - 1} \\
    & \qquad \cross x_4^{\omega(G \backslash \{e_1, e_2, e_3\}) - \omega(G \backslash \{e_1, 
    e_2, e_3, 
    e_4 \}) - 1} f(x(\xi)) \\ 
    &= \frac{1}{I^{\tr}} \frac{1}{\omega(G \backslash \{e_1, e_2, e_3,  e_4 \}}
    \frac{1}{\omega(G \backslash \{e_1, e_2, e_3\}} 
    \int_0^{1} \dd{x_2} 
    \int_0^{x_2} \dd{x_3} 
    \int_{0}^{1} \dd{\xi_4} \int_0^{1} \dd{\xi_5} \\ & \qquad \cross x_2^{ \omega(G \backslash 
    \{e_1\}) - \omega(G \backslash \{e_1, e_2\}) - \omega(G \backslash \{e_1, 
    e_2, e_3\}) 
    - \omega(G \backslash \{e_1, e_2, e_3, 
    e_4 \}) - 1} \\ & \qquad \cross  x_3^{\omega(G \backslash \{e_1, e_2\}) - \omega(G 
    \backslash 
    \{e_1, e_2, e_3\}) - \omega(G \backslash \{e_1, e_2, e_3, 
    e_4 \}) - 1} f(x(\xi)) \\ 
    &= \frac{1}{I^{\tr}} \frac{1}{\omega(G \backslash \{e_1, e_2, e_3,  e_4 \}}
    \frac{1}{\omega(G \backslash \{e_1, e_2, e_3\}} 
    \frac{1}{\omega(G \backslash \{e_1, e_2\}} 
    \int_0^{1} \dd{x_2} 
    \int_0^{1} \dd{\xi_3} 
    \int_{0}^{1} \dd{\xi_4} \int_0^{1} \dd{\xi_5} \\ & \qquad \cross x_2^{ \omega(G \backslash 
    \{e_1\}) - \omega(G \backslash \{e_1, e_2\}) - \omega(G \backslash \{e_1, 
    e_2, e_3\}) 
    - \omega(G \backslash \{e_1, e_2, e_3, 
    e_4 \}) - 1} f(x(\xi)) \\ 
    &= \frac{1}{I_{tr}} \frac{1}{\omega(G \backslash \{e_1, e_2, e_3,  e_4 \}}
    \frac{1}{\omega(G \backslash \{e_1, e_2, e_3\}} 
    \frac{1}{\omega(G \backslash \{e_1, e_2\}} 
    \frac{1}{\omega(G \backslash \{e_1\}} \\
    & \qquad \cross \int_0^{1} \dd{\xi_2} 
    \int_0^{1} \dd{\xi_3} 
    \int_{0}^{1} \dd{\xi_4} \int_0^{1} \dd{\xi_5} \; f(x(\xi)) \;.
\end{split}
\end{equation*}
We can thus sample a set of Feynman parameters by first sampling the variables $\xi_e$ from the unit interval, and then plugging these values into the coordinate transformation~\eqref{eq:transform}.

The remaining steps of Section~\ref{sec:samp} take these sampled Feynman parameters and use them to sample loop momenta. For completeness, we give explicit expressions for the required objects.

We route the momenta such that the loop-edge momentum matrix $\mathcal{M}$ is given by
\begin{equation*}
    \mathcal{M} = \begin{pmatrix}
        1 & 0 \\
        1 & 0 \\ 
        1 & -1 \\ 
        0 & 1 \\ 
        0 & 1 \\
    \end{pmatrix}
\end{equation*}
and such that $p_2 = p_5 = p$. We can compute the matrix $\mathcal{L}$ from $\mathcal{M}$: 
\begin{equation*}
    \mathcal{L} = \sum_e x_e \mathcal{M}_{e, l} \mathcal{M}_{e, l'} = 
    \begin{pmatrix}
        x_1 + x_2 + x_3 & -x_3 \\
        -x_3 & x_3 + x_4 + x_5 \\
    \end{pmatrix}
\end{equation*}
From this, we verify that: 
\begin{equation*}
\begin{split}
    \det(\mathcal{L}) &= (x_1 + x_2 + x_3) (x_3 + x_4 + x_5) - x_3^2 \\ &=
    x_1 x_3 + x_1 x_4 + x_1 x_5 + x_2 x_3 + x_2 x_4 + x_2 x_5 + x_3 x_4 
    + x_3 x_5 \\ 
    &= \mathcal{U}(x)
\end{split}
\end{equation*}
The inverse matrix is given by
\begin{equation*}
    \mathcal{L}^{-1} = \frac{1}{\mathcal{U}(x)} 
    \begin{pmatrix}
        x_3 + x_4 + x_5 & x_3 \\ 
        x_3 & x_1 + x_2 + x_3
    \end{pmatrix}
\end{equation*}
A Cholesky decomposition of $\mathcal{L}$ is 
\begin{equation*}
\begin{split}
    Q &= \begin{pmatrix}
        \sqrt{x_1 + x_2 + x_3} & \frac{-x_3}{\sqrt{x_1 + x_2 + x_3}} \\
        0 & \sqrt{x_3 + x_4 + x_5 - \frac{x_3^2}{x_1 + x_2 + x_3}}
    \end{pmatrix}  \\ 
    &= \begin{pmatrix}
        \sqrt{x_1 + x_2 + x_3} & \frac{-x_3}{\sqrt{x_1 + x_2 + x_3}} \\
        0 & \sqrt{\frac{\mathcal{U}(x)}{x_1 + x_2 + x_3}} \;,
    \end{pmatrix}
\end{split}
\end{equation*}
which has the inverse:
\begin{equation*}
    Q^{-1} = \frac{1}{\sqrt{\mathcal{U}(x)}}
    \begin{pmatrix}
       \sqrt{\frac{\mathcal{U}(x)}{x_1 + x_2 + x_3}} & \frac{-x_3}{\sqrt{x_1 + x_2 + x_3}} \\
        0 &  \sqrt{x_1 + x_2 + x_3}  \;.
    \end{pmatrix}  \\ 
\end{equation*}
We also need the vector $\mathbf{u}$, which in this example is given by: 
\begin{equation*}
    \mathbf{u} = \begin{pmatrix}
        x_2 p \\ 
        x_5 p
    \end{pmatrix}    
\end{equation*}
After sampling $\lambda$ from the gamma distribution, we have all the ingredients to construct a set of loop momenta using Eq.~\eqref{eq:loop_momenta}.


\begin{thebibliography}{10}
\expandafter\ifx\csname url\endcsname\relax
  \def\url#1{\texttt{#1}}\fi
\expandafter\ifx\csname urlprefix\endcsname\relax\def\urlprefix{URL }\fi
\expandafter\ifx\csname href\endcsname\relax
  \def\href#1#2{#2} \def\path#1{#1}\fi

\bibitem{Kleiss:1985gy}
R.~Kleiss, W.~J. Stirling, S.~D. Ellis, A new {Monte Carlo} treatment of
  multiparticle phase space at high-energies, Comput. Phys. Commun. 40 (1986)
  359.
\newblock \href {https://doi.org/10.1016/0010-4655(86)90119-0}
  {\path{doi:10.1016/0010-4655(86)90119-0}}.

\bibitem{Papadopoulos:2000tt}
C.~G. Papadopoulos, {PHEGAS: A Phase space generator for automatic
  cross-section computation}, Comput. Phys. Commun. 137 (2001) 247--254.
\newblock \href {http://arxiv.org/abs/hep-ph/0007335}
  {\path{arXiv:hep-ph/0007335}}, \href
  {https://doi.org/10.1016/S0010-4655(01)00163-1}
  {\path{doi:10.1016/S0010-4655(01)00163-1}}.

\bibitem{Gehrmann-DeRidder:2003pne}
A.~Gehrmann-De~Ridder, T.~Gehrmann, G.~Heinrich, {Four particle phase space
  integrals in massless QCD}, Nucl. Phys. B 682 (2004) 265--288.
\newblock \href {http://arxiv.org/abs/hep-ph/0311276}
  {\path{arXiv:hep-ph/0311276}}, \href
  {https://doi.org/10.1016/j.nuclphysb.2004.01.023}
  {\path{doi:10.1016/j.nuclphysb.2004.01.023}}.

\bibitem{Binoth:2004jv}
T.~Binoth, G.~Heinrich, {Numerical evaluation of phase space integrals by
  sector decomposition}, Nucl. Phys. B 693 (2004) 134--148.
\newblock \href {http://arxiv.org/abs/hep-ph/0402265}
  {\path{arXiv:hep-ph/0402265}}, \href
  {https://doi.org/10.1016/j.nuclphysb.2004.06.005}
  {\path{doi:10.1016/j.nuclphysb.2004.06.005}}.

\bibitem{Borinsky:2020rqs}
M.~Borinsky, {Tropical Monte Carlo quadrature for Feynman integrals}, Ann.
  Inst. H. Poincare D Comb. Phys. Interact. 10~(4) (2023) 635--685.
\newblock \href {http://arxiv.org/abs/2008.12310} {\path{arXiv:2008.12310}},
  \href {https://doi.org/10.4171/aihpd/158} {\path{doi:10.4171/aihpd/158}}.

\bibitem{Borinsky:2023jdv}
M.~Borinsky, H.~J. Munch, F.~Tellander, {Tropical Feynman integration in the
  Minkowski regime}, Comput. Phys. Commun. 292 (2023) 108874.
\newblock \href {http://arxiv.org/abs/2302.08955} {\path{arXiv:2302.08955}},
  \href {https://doi.org/10.1016/j.cpc.2023.108874}
  {\path{doi:10.1016/j.cpc.2023.108874}}.

\bibitem{Binoth:2000ps}
T.~Binoth, G.~Heinrich, {An automatized algorithm to compute infrared divergent
  multiloop integrals}, Nucl. Phys. B 585 (2000) 741--759.
\newblock \href {http://arxiv.org/abs/hep-ph/0004013}
  {\path{arXiv:hep-ph/0004013}}, \href
  {https://doi.org/10.1016/S0550-3213(00)00429-6}
  {\path{doi:10.1016/S0550-3213(00)00429-6}}.

\bibitem{Kaneko:2009qx}
T.~Kaneko, T.~Ueda, {A Geometric method of sector decomposition}, Comput. Phys.
  Commun. 181 (2010) 1352--1361.
\newblock \href {http://arxiv.org/abs/0908.2897} {\path{arXiv:0908.2897}},
  \href {https://doi.org/10.1016/j.cpc.2010.04.001}
  {\path{doi:10.1016/j.cpc.2010.04.001}}.

\bibitem{Panzer:2019yxl}
E.~Panzer, {Hepps bound for Feynman graphs and matroids}, Ann. Inst. H.
  Poincare D Comb. Phys. Interact. 10~(1) (2022) 31--119.
\newblock \href {http://arxiv.org/abs/1908.09820} {\path{arXiv:1908.09820}},
  \href {https://doi.org/10.4171/aihpd/126} {\path{doi:10.4171/aihpd/126}}.

\bibitem{Brown:2015fyf}
F.~Brown, {Feynman amplitudes, coaction principle, and cosmic Galois group},
  Commun. Num. Theor. Phys. 11 (2017) 453--556.
\newblock \href {http://arxiv.org/abs/1512.06409} {\path{arXiv:1512.06409}},
  \href {https://doi.org/10.4310/CNTP.2017.v11.n3.a1}
  {\path{doi:10.4310/CNTP.2017.v11.n3.a1}}.

\bibitem{Catani:2008xa}
S.~Catani, T.~Gleisberg, F.~Krauss, G.~Rodrigo, J.-C. Winter, {From loops to
  trees by-passing Feynman's theorem}, JHEP 09 (2008) 065.
\newblock \href {http://arxiv.org/abs/0804.3170} {\path{arXiv:0804.3170}},
  \href {https://doi.org/10.1088/1126-6708/2008/09/065}
  {\path{doi:10.1088/1126-6708/2008/09/065}}.

\bibitem{Bierenbaum:2010cy}
I.~Bierenbaum, S.~Catani, P.~Draggiotis, G.~Rodrigo, A tree-loop duality
  relation at two loops and beyond, JHEP 10 (2010) 073.
\newblock \href {http://arxiv.org/abs/1007.0194} {\path{arXiv:1007.0194}},
  \href {https://doi.org/10.1007/JHEP10(2010)073}
  {\path{doi:10.1007/JHEP10(2010)073}}.

\bibitem{Lepage:1977sw}
G.~P. Lepage, A new algorithm for adaptive multidimensional integration, J.
  Comput. Phys. 27 (1978) 192.
\newblock \href {https://doi.org/10.1016/0021-9991(78)90004-9}
  {\path{doi:10.1016/0021-9991(78)90004-9}}.

\bibitem{Kleiss:1994qy}
R.~Kleiss, R.~Pittau, {Weight optimization in multichannel Monte Carlo},
  Comput. Phys. Commun. 83 (1994) 141--146.
\newblock \href {http://arxiv.org/abs/hep-ph/9405257}
  {\path{arXiv:hep-ph/9405257}}, \href
  {https://doi.org/10.1016/0010-4655(94)90043-4}
  {\path{doi:10.1016/0010-4655(94)90043-4}}.

\bibitem{Heimel:2022wyj}
T.~Heimel, R.~Winterhalder, A.~Butter, J.~Isaacson, C.~Krause, F.~Maltoni,
  O.~Mattelaer, T.~Plehn, {MadNIS - Neural multi-channel importance sampling},
  SciPost Phys. 15~(4) (2023) 141.
\newblock \href {http://arxiv.org/abs/2212.06172} {\path{arXiv:2212.06172}},
  \href {https://doi.org/10.21468/SciPostPhys.15.4.141}
  {\path{doi:10.21468/SciPostPhys.15.4.141}}.

\bibitem{andrea}
A.~Favorito, Tropical {Feynman} period integration, {Master's Thesis (Available
  at \url{https://michaelborinsky.com/static/thesis_favorito.pdf})}, ETH
  Z\"urich, Z\"urich, Switzerland (March 2023).

\bibitem{Beekveldt:2020kzk}
R.~Beekveldt, M.~Borinsky, F.~Herzog, {The Hopf algebra structure of the
  $R^\star$-operation}, JHEP 07 (2020) 061.
\newblock \href {http://arxiv.org/abs/2003.04301} {\path{arXiv:2003.04301}},
  \href {https://doi.org/10.1007/JHEP07(2020)061}
  {\path{doi:10.1007/JHEP07(2020)061}}.

\bibitem{10.5555/270146}
D.~E. Knuth, The Art of Computer Programming, 3rd Edition, Vol.~2,
  Addison-Wesley.

\bibitem{Capatti:2020xjc}
Z.~Capatti, V.~Hirschi, A.~Pelloni, B.~Ruijl, {Local unitarity: a
  representation of differential cross-sections that is locally free of
  infrared singularities at any order}, JHEP 04 (2021) 104.
\newblock \href {http://arxiv.org/abs/2010.01068} {\path{arXiv:2010.01068}},
  \href {https://doi.org/10.1007/JHEP04(2021)104}
  {\path{doi:10.1007/JHEP04(2021)104}}.

\bibitem{Becker:2012aqa}
S.~Becker, C.~Reuschle, S.~Weinzierl, Efficiency improvements for the numerical
  computation of nlo corrections, JHEP 07 (2012) 090.
\newblock \href {http://arxiv.org/abs/1205.2096} {\path{arXiv:1205.2096}},
  \href {https://doi.org/10.1007/JHEP07(2012)090}
  {\path{doi:10.1007/JHEP07(2012)090}}.

\bibitem{Capatti:2022tit}
Z.~Capatti, V.~Hirschi, B.~Ruijl, {Local unitarity: cutting raised propagators
  and localising renormalisation}, JHEP 10 (2022) 120.
\newblock \href {http://arxiv.org/abs/2203.11038} {\path{arXiv:2203.11038}},
  \href {https://doi.org/10.1007/JHEP10(2022)120}
  {\path{doi:10.1007/JHEP10(2022)120}}.

\bibitem{incomplete_gamma}
A.~R. DiDonato, A.~H. Morris,
  \href{https://doi.org/10.1145/22721.23109}{Computation of the incomplete
  gamma function ratios and their inverse}, ACM Trans. Math. Softw. 12~(4)
  (1986) 377–393.
\newblock \href {https://doi.org/10.1145/22721.23109}
  {\path{doi:10.1145/22721.23109}}.
\newline\urlprefix\url{https://doi.org/10.1145/22721.23109}

\bibitem{Sborlini:2016gbr}
G.~F.~R. Sborlini, F.~Driencourt-Mangin, R.~Hernandez-Pinto, G.~Rodrigo,
  {Four-dimensional unsubtraction from the loop-tree duality}, JHEP 08 (2016)
  160.
\newblock \href {http://arxiv.org/abs/1604.06699} {\path{arXiv:1604.06699}},
  \href {https://doi.org/10.1007/JHEP08(2016)160}
  {\path{doi:10.1007/JHEP08(2016)160}}.

\bibitem{Sborlini:2016hat}
G.~F.~R. Sborlini, F.~Driencourt-Mangin, G.~Rodrigo, {Four-dimensional
  unsubtraction with massive particles}, JHEP 10 (2016) 162.
\newblock \href {http://arxiv.org/abs/1608.01584} {\path{arXiv:1608.01584}},
  \href {https://doi.org/10.1007/JHEP10(2016)162}
  {\path{doi:10.1007/JHEP10(2016)162}}.

\bibitem{Capatti:2019ypt}
Z.~Capatti, V.~Hirschi, D.~Kermanschah, B.~Ruijl, Loop-tree duality for
  multiloop numerical integration, Phys. Rev. Lett. 123~(15) (2019) 151602.
\newblock \href {http://arxiv.org/abs/1906.06138} {\path{arXiv:1906.06138}},
  \href {https://doi.org/10.1103/PhysRevLett.123.151602}
  {\path{doi:10.1103/PhysRevLett.123.151602}}.

\bibitem{Capatti:2019edf}
Z.~Capatti, V.~Hirschi, D.~Kermanschah, A.~Pelloni, B.~Ruijl, Numerical
  loop-tree duality: contour deformation and subtraction, JHEP 04 (2020) 096.
\newblock \href {http://arxiv.org/abs/1912.09291} {\path{arXiv:1912.09291}},
  \href {https://doi.org/10.1007/JHEP04(2020)096}
  {\path{doi:10.1007/JHEP04(2020)096}}.

\bibitem{deLejarza:2024scm}
J.~J.~M. de~Lejarza, D.~F. Renter\'\i{}a-Estrada, M.~Grossi, G.~Rodrigo,
  {Quantum integration of decay rates at second order in perturbation theory},
  Quantum Sci. Technol. 10~(2) (2025) 025026.
\newblock \href {http://arxiv.org/abs/2409.12236} {\path{arXiv:2409.12236}},
  \href {https://doi.org/10.1088/2058-9565/ada9c5}
  {\path{doi:10.1088/2058-9565/ada9c5}}.

\bibitem{Soper:1999xk}
D.~E. Soper, {Techniques for QCD calculations by numerical integration}, Phys.
  Rev. D 62 (2000) 014009.
\newblock \href {http://arxiv.org/abs/hep-ph/9910292}
  {\path{arXiv:hep-ph/9910292}}, \href
  {https://doi.org/10.1103/PhysRevD.62.014009}
  {\path{doi:10.1103/PhysRevD.62.014009}}.

\bibitem{Gong:2008ww}
W.~Gong, Z.~Nagy, D.~E. Soper, {Direct numerical integration of one-loop
  Feynman diagrams for N-photon amplitudes}, Phys. Rev. D 79 (2009) 033005.
\newblock \href {http://arxiv.org/abs/0812.3686} {\path{arXiv:0812.3686}},
  \href {https://doi.org/10.1103/PhysRevD.79.033005}
  {\path{doi:10.1103/PhysRevD.79.033005}}.

\bibitem{AH:2023kor}
A.~A~H, E.~Chaubey, M.~Fraaije, V.~Hirschi, H.-S. Shao, {Light-by-light
  scattering at next-to-leading order in QCD and QED}, Phys. Lett. B 851 (2024)
  138555.
\newblock \href {http://arxiv.org/abs/2312.16956} {\path{arXiv:2312.16956}},
  \href {https://doi.org/10.1016/j.physletb.2024.138555}
  {\path{doi:10.1016/j.physletb.2024.138555}}.

\bibitem{Navarrete:2024zgz}
P.~Navarrete, R.~Paatelainen, K.~Sepp\"anen, {Perturbative QCD meets phase
  quenching: The pressure of cold quark matter}, Phys. Rev. D 110~(9) (2024)
  094033.
\newblock \href {http://arxiv.org/abs/2403.02180} {\path{arXiv:2403.02180}},
  \href {https://doi.org/10.1103/PhysRevD.110.094033}
  {\path{doi:10.1103/PhysRevD.110.094033}}.

\bibitem{Kermanschah:2024utt}
D.~Kermanschah, M.~Vicini, {$N_f$-contribution to the virtual correction for
  electroweak vector boson production at NNLO} (7 2024).
\newblock \href {http://arxiv.org/abs/2407.18051} {\path{arXiv:2407.18051}}.

\bibitem{Capatti:2020ytd}
Z.~Capatti, V.~Hirschi, D.~Kermanschah, A.~Pelloni, B.~Ruijl, Manifestly causal
  loop-tree duality (2020).
\newblock \href {http://arxiv.org/abs/2009.05509} {\path{arXiv:2009.05509}}.

\bibitem{Capatti:2022mly}
Z.~Capatti, {Exposing the threshold structure of loop integrals}, Phys. Rev. D
  107~(5) (2023) L051902.
\newblock \href {http://arxiv.org/abs/2211.09653} {\path{arXiv:2211.09653}},
  \href {https://doi.org/10.1103/PhysRevD.107.L051902}
  {\path{doi:10.1103/PhysRevD.107.L051902}}.

\bibitem{vanHameren:2010cp}
A.~van Hameren, {OneLOop: For the evaluation of one-loop scalar functions},
  Comput. Phys. Commun. 182 (2011) 2427--2438.
\newblock \href {http://arxiv.org/abs/1007.4716} {\path{arXiv:1007.4716}},
  \href {https://doi.org/10.1016/j.cpc.2011.06.011}
  {\path{doi:10.1016/j.cpc.2011.06.011}}.

\bibitem{Borowka:2017idc}
S.~Borowka, G.~Heinrich, S.~Jahn, S.~P. Jones, M.~Kerner, J.~Schlenk, T.~Zirke,
  {pySecDec: a toolbox for the numerical evaluation of multi-scale integrals},
  Comput. Phys. Commun. 222 (2018) 313--326.
\newblock \href {http://arxiv.org/abs/1703.09692} {\path{arXiv:1703.09692}},
  \href {https://doi.org/10.1016/j.cpc.2017.09.015}
  {\path{doi:10.1016/j.cpc.2017.09.015}}.

\bibitem{Ruijl:2017cxj}
B.~Ruijl, T.~Ueda, J.~A.~M. Vermaseren, {Forcer, a FORM program for the
  parametric reduction of four-loop massless propagator diagrams}, Comput.
  Phys. Commun. 253 (2020) 107198.
\newblock \href {http://arxiv.org/abs/1704.06650} {\path{arXiv:1704.06650}},
  \href {https://doi.org/10.1016/j.cpc.2020.107198}
  {\path{doi:10.1016/j.cpc.2020.107198}}.

\bibitem{Basso:2017jwq}
B.~Basso, L.~J. Dixon, {Gluing Ladder Feynman Diagrams into Fishnets}, Phys.
  Rev. Lett. 119~(7) (2017) 071601.
\newblock \href {http://arxiv.org/abs/1705.03545} {\path{arXiv:1705.03545}},
  \href {https://doi.org/10.1103/PhysRevLett.119.071601}
  {\path{doi:10.1103/PhysRevLett.119.071601}}.

\bibitem{Hirschi:2011pa}
V.~Hirschi, R.~Frederix, S.~Frixione, M.~V. Garzelli, F.~Maltoni, R.~Pittau,
  {Automation of one-loop QCD corrections}, JHEP 05 (2011) 044.
\newblock \href {http://arxiv.org/abs/1103.0621} {\path{arXiv:1103.0621}},
  \href {https://doi.org/10.1007/JHEP05(2011)044}
  {\path{doi:10.1007/JHEP05(2011)044}}.

\bibitem{Balduf:2023ilc}
P.-H. Balduf, {Statistics of Feynman amplitudes in
  {\ensuremath{\phi}}$^{4}$-theory}, JHEP 11 (2023) 160.
\newblock \href {http://arxiv.org/abs/2305.13506} {\path{arXiv:2305.13506}},
  \href {https://doi.org/10.1007/JHEP11(2023)160}
  {\path{doi:10.1007/JHEP11(2023)160}}.

\bibitem{Borinsky:2025ywo}
M.~Borinsky, A.~Favorito, {Feynman integrals at large loop order and the
  $\log$-$\Gamma$ distribution} (2025).
\newblock \href {http://arxiv.org/abs/2503.07803} {\path{arXiv:2503.07803}}.

\end{thebibliography}
\end{document}